\documentclass[12pt,preprint]{aastex}

\newcommand{\ud}[2]{\mbox{$^{+ #1}_{- #2}$}}

\def \be{\begin{eqnarray}}
\def \ee{\end{eqnarray}}

\received{}
\revised{}
\accepted{}
\shorttitle{RX~J185635-3754}
\shortauthors{Pons et al.}

\begin{document}

\title{TOWARDS A MASS AND RADIUS DETERMINATION OF 
THE NEARBY ISOLATED NEUTRON STAR RX~J185635-3754}
\author{ JOSE A. PONS$^{1,2}$,  FREDERICK M. WALTER$^1$, JAMES M. LATTIMER$^1$, \\
MADAPPA PRAKASH$^1$, RALPH NEUH\"AUSER$^3$, AND PENGHUI AN$^1$ }
\affil{$^1$Department of Physics \& Astronomy,
      	State University of New York at Stony Brook, \\
       	Stony Brook, NY 11794-3800, USA}
\email{jose.pons@roma1.infn.it, fwalter@astro.sunysb.edu,
       lattimer@astro.sunysb.edu, prakash@nuclear.physics.sunysb.edu,
       pan@astro.sunysb.edu}
 
\affil{$^2$ Dipartimento di Fisica ``G.Marconi'', Universita' di Roma
        ``La Sapienza'', 00185 Roma, Italy}
\affil{$^3$ MPI f\"ur extraterrestrische Physik, 85740 Garching, Germany}
\email{rne@xray.mpe.mpg.de}

\begin{abstract}

We discuss efforts to determine the mass, radius, and surface composition
of the nearby compact object RX~J185635-3754 from its multi-wavelength
spectral energy distribution. We compute non-magnetized model
atmospheres and emergent spectra for selected compositions and
gravities, and discuss efforts to fit existing and new observational
data from {\it ROSAT, EUVE} and the {\it HST}.
The spectral energy
distribution matches that expected from a heavy-element
dominated atmosphere, but not from a uniform temperature blackbody.
Non-magnetic light element atmospheres cannot be simultaneously 
reconciled with the optical and X-ray data.
We extend previous studies, which were limited to one fixed
neutron star mass and radius. For uniform temperature models dominated by 
heavy elements, the redshift $z$ is constrained to be
0.3$\lesssim z \lesssim$~0.4 and
the best-fit mass and radius are $M\approx 0.9$~M$_\odot$ 
and $R\approx 6$ km (for a 61~pc distance).
These values for $M$ and $R$ together are not permitted for any plausible 
equation of state,
including that of a self-bound strange quark star.
A simplified two-temperature model 
allows masses and radii up to about 50\% larger,
or a factor of 2 in the case of a black body. 
The observed luminosity is consistent with the thermal emission of an isolated
neutron star no older than about 1 million years, the age inferred from
available proper motion and parallax information. 
   
\end{abstract}

\keywords{dense matter --- stars: neutron 
--- stars: fundamental parameters --- stars: individual (RX J185635-3754)}

\newpage

\section{INTRODUCTION}

It has long been hoped that stringent constraints could be placed on the
equation of state (EOS) of dense matter from astrophysical measurements.
 Up to the present time, however, astrophysical constraints, most
important of which is the establishment of a reliable minimum value for
the neutron star's maximum mass (1.44 M$_\odot$ from the binary pulsar
PSR 1913+16 \citep{Tho94}), have not proven enormously useful in
limiting the dense matter EOS. Although accurate masses of several
additional neutron stars are available \citep{TC99}, a precise
measurement of the radius does not yet exist. There are observational
estimates of radii, and limits to both mass and radius, 
such as those from Quasi-Periodic Oscillators (QPOs),
X-ray bursts, and pulsars (see \cite{LP2000} for a review).  Even less
information concerning the atmospheric composition is known.  
Spectra of neutron stars and pulsars have been fit with blackbody
spectra (e.g., Paerels et al.\/ 2000) 
and pure hydrogen atmospheres (e.g.,
Rutledge et al.\/ 2001).    
The purpose of this paper is to present estimates of the mass,
radius, and atmospheric composition of the nearby compact object,
RX~J185635-3754 based on model fitting to multiwavelength data.
In spite of the lack of specific observed spectral
features, we find that the star's redshift is  constrained.  
The current generation of X-ray satellites will soon permit direct measurements
of the gravitational redshift and surface  composition.

\cite{WWN96} discovered the X-ray source RX~J185635-3754,
and showed that its emergent flux was consistent with  a
blackbody (BB) spectrum with an effective temperature  (observed at the
Earth) of $T_\infty\simeq57$ eV.  The fortuitous location of the source
in the foreground of the R~CrA molecular cloud provides an upper limit
to the distance $D<130$ pc.  The observed total flux
implies that, for uniform surface temperatures, the so-called radiation
radius $R_\infty=R/\sqrt{1-2GM/Rc^2}$ is only a few km.  On this basis,
they concluded it must be an isolated neutron star. Using the Hubble
Space Telescope ({\it HST}), \cite{WM97} subsequently identified a source at
optical (6060 \AA) and near-ultraviolet (3000 \AA) wavelengths,
about 2.5 and 4 times brighter,
respectively, than an extrapolation of a 57 eV BB
into these bands.  \cite{Neu98} detected the object with $V$=25.7 using
the 3.6 meter New Technology Telescope ({\it NTT}).
Both the ultraviolet and red magnitudes have been
confirmed by the subsequent {\it HST} measurements reported below.  X-ray
observations of RX~J185635-3754, at energies greater
than 1 keV, failed to show any evidence of a non-thermal tail or
cyclotron emission lines. There is no reported detection of a radio
counterpart, and Walter et al.\/ (1996)
reported a limit to any X-ray variability of 7\%.  
The optical flux rules out any active accretion disk or the presence
of a binary companion with a stellar spectrum  more luminous than about 
10$^{-7}$~L$_\odot$ (in the optical band).
We interpret these results to mean that there is neither
significant accretion nor an intense magnetosphere. RX~J185635-3754
appears to be a radio-quiet, isolated neutron star.

Three sets of observations with the {\it HST} Wide Field Planetary Camera 2
(WFPC2), taken within an
interval of about three years, have yielded the proper motion ($332
\pm 1$ mas~yr$^{-1}$) and parallax ($16.5 \pm 2.3$ mas), which
corresponds, respectively, to a transverse velocity of about
$108\pm15$ km~s$^{-1}$ and $D=61^{+9}_{-8}$ pc \citep{Wal01}.
This velocity and distance
imply that the star originated $9\pm2\times$10$^5$ years ago in
the Upper Scorpius OB association \citep{Wal01}.  This
interpretation is supported by a possible association with the runaway
OB star $\zeta$ Oph, which also seems to have originated in the Upper
Sco association at a similar time.

A space velocity greater than 10 km~s$^{-1}$ virtually rules out the
possibility that the neutron star has significant accretion
\citep{MB94}.  Unlike some other radio-quiet isolated
neutron stars \citep{Car96}, which have pronounced evidence of
non-thermal emission, this object may provide a clear view of the
surface of a neutron star without complications from a magnetosphere
or accretion.  Ultimately, it holds the prospect that both its radius
and mass, and the EOS of dense matter, might be tightly
constrained.

The paper is organized as follows:  The existing and new data
available are presented in \S 2.  In \S 3 we present a timing
analysis and put limits on the modulation of the X-ray flux.  In \S 4, we
construct non-magnetic neutron star atmosphere models for selected
chemical compositions and compare our results to previous studies.  In
\S 5, fits to the multiwavelength spectral energy distribution
are computed, assuming that
the neutron star's surface temperature is uniform. In \S 6, we
explore the consequences of non-uniform thermal emission.  Finally, in
\S 7, we discuss some
implications of our results, focusing on other aspects of the star
and outlining other theoretical
issues that might affect our interpretations.

\section{SUMMARY OF OBSERVATIONS}

\subsection{X-ray Observations}

\subsubsection{ROSAT}

{\it ROSAT} observed this target on three occasions
(Table~\ref{tbl-rosobs}). The Position Sensitive Proportional Counter
(PSPC) observation and the first of the two
High Resolution Imager (HRI)
observations were discussed by Walter et al.\/ (1996).
The second HRI
observation was obtained 3 years after the first for the purpose of
studying the variability and proper motion of the target.

The HRI source counts were extracted from within a circle of radius
72~arcsec centered on the source centroid; the background was taken
from a concentric annulus of inner and outer radii 72 and 216~arcsec,
respectively.  The net count rates were 0.555$\pm$0.006 and
0.556$\pm$0.005 counts~s$^{-1}$ for the two observations; the source thus shows
no evidence of long-term variability.  Comparison of the position of
this X-ray source with respect to 6 other X-ray sources in the HRI
observations indicated that the target has a proper motion of $400\pm200$
milli-arcsec towards the south--east. This is consistent with the
superior {\it HST} measurement \citep{Wal01}.

\subsubsection{EUVE}

The Extreme UltraViolet Explorer ({\it EUVE}; Haisch, Bowyer, \& Malina 1993)
pointed at the target for the better part of a month in June/July
1997 (Table~\ref{tbl-euve}), for a total of 582.6 ks. 
The target was clearly
detected in the Deep Sky Survey (DSS) imager and in the Short Wavelength
(SW) Spectrograph.

The DSS Image was taken in WSZ mode, which allows for pulse-height
discrimination. Following the prescription of \cite{Chr95}, we
filtered the counts, selecting only those with pulse heights between
8500 and 15,500.  This resulted in a 55\%
decrease in background counts with a loss of 14\% of the source
counts. 
The signal-to-noise ratio is commensurately
increased by this pulse-height filtering.

We edited the list of 335 good time intervals, removing those intervals
during which no counts were detected (there were 10 such intervals).  At
the beginning and end of each interval, we trimmed any time without
photons provided the length of this period exceeded three times the
expected time between counts, and also eliminated the frequent 2-3
minute period (extending up to 20 minutes during a few intervals) at the
end of the observation without any counts. Doing so decreased the good
exposure time by 14\% to 499~ks.

The extracted pulse-height filtered data (mostly background) have a mean
rate of about 1 count~s$^{-1}$. We examined the instantaneous count
rates (using a 20 second running mean) and eliminated all intervals
where the total count rate was zero, or greater than 2~counts~s$^{-1}$. The
former represent times when the detector was not turned on; the latter
are contaminated by scattered solar radiation from the limb of the
Earth, or are contaminated by the particle radiation in the
South Atlantic Anomaly.
Following this filtering the net observing time is 478,696 s.

The corrected DSS count rate is 0.0429$\pm$0.0003 counts~s$^{-1}$. Within the
uncertainties, this is consistent with the 3$\sigma$ upper limit of
$<$0.066~counts~s$^{-1}$ reported by Walter et al.\/ (1996),
and with the detection at
0.028 counts~s$^{-1}$ reported by \cite{Lam97}. Both of those observations were
made  through the Lexan scanner filter, which is thicker and has lower
throughput than the DSS imager. No other sources were detected in the
DSS image.

To obtain the {\it EUVE} spectrum we created two-dimensional images in
wavelength-$\theta$ space  ($\theta$ is the angle off-axis perpendicular
to the dispersion axis) from the pulse-height filtered data for each of
the spectrometers.  The dispersed spectrum was clearly present in the 
SW detector, with an excess of counts at $\theta=0$; as expected, there
is no detection in the Medium or Long Wavelength detectors. 
We summed the counts in the SW image
over angles $\theta=0\pm0.003^o$. We summed the background counts
from $0.004^o-0.024^o$ above and below the spectrum, and smoothed them
with a 3\AA\ running boxcar prior to subtraction. We truncated the
spectrum 
shortward of 76 \AA.  We divided the counts spectrum by the effective
area of the SW detector (sw\_ea.tab version dated 6 March 1997) and by
the net exposure time, and corrected for the 14\% signal loss in the
pulse-height filtering. The net spectrum is shown in
Figure~\ref{fig_euve}.
Uncertainties are computed from counting statistics.  A
continuum is clearly detected between about 75 and 95~\AA, with net flux
of 2.5$\pm$0.3$\times$10$^{-13}$ erg cm$^{-2}$ s$^{-1}$ in this 20~\AA~
interval.


\subsubsection{ASCA}

We observed the target with the Advanced Satellite for Cosmology and
Astrophysics ({\it ASCA}) beginning on 1997
October 11 at 15:26~UT and ending at 13:34~UT on October 12 (sequence
25037000). The Solid-state Imaging Spectrometer (SIS) exposure time
is 36.6~ks; the  Gas Imaging Spectrometer (GIS)
exposure time is 41.4~ks. We observed with a single
CCD in bright mode.

The target was detected in the two SIS detectors, with 480$\pm$33
counts. There is a marginal detection in the lowest channels of the
GIS.  In neither case is there any detection at energies above
1.5~keV. We extracted the spectra using both FTOOLS and software
written in IDL.
The background is the mean of background regions extracted from around the
source. The net SIS spectrum is shown in Figure~\ref{fig_asca}.

We fit the SIS spectrum to a BB using Version 10 of XSPEC. The best-fit
BB parameters ($T_\infty=60\pm^{20}_{25}$~eV, $n_{H,20}<70$) are
consistent with, though less well-determined than, the {\it ROSAT} PSPC fit
(Walter et al.\/ 1996).
The total observed flux is $2.2\times10^{-12}$
erg~cm$^{-2}$~s$^{-1}$.  The low count rate did not warrant more
detailed spectral fitting.  We show the net GIS spectrum in Figure
\ref{fig_asca}. Given the low S/N of the detection and the uncertain
response of the GIS below 1 keV, we did not attempt to fit these data,
other than to note that the data are consistent with a soft BB spectrum.
Consequently, we do not use these data in the multispectral fits.

There is no evidence for any line emission at energies above 1.5~keV.
Such line emission might be expected
from a magnetized neutron star accreting from the interstellar medium
\citep{Nel95}. Neither is there any evidence for a hard tail, as would be
expected in an accreting source due to comptonization of the infalling
electrons \citep{Zam95}. The $2\sigma$ upper limits on the 1.5--10~keV
flux are $1.01\times10^{-5}$~photons~cm$^{-2}$~s$^{-1}$ in the SIS
and $3.8\times10^{-5}$~photons~cm$^{-2}$~s$^{-1}$ in the GIS.

No other X-ray sources were detected in the SIS.
The source RX~J185533-3805 was detected in the GIS (it is outside the SIS
field of view).  Its spectrum can be fit either with a 700~eV
BB or an $\alpha$=1.9 power law. The nature of this object is not
known; there is no obvious optical counterpart.

\subsection{UV/Optical Photometry}

\subsubsection{Broadband Imaging from the HST}
We observed and detected the target with the WFPC2 
using the F606W, F450W, F303W, and F178W filters (Table \ref{tbl-hst}). 
The three
F606W images have been used to determine the parallax and proper motion
of the target \citep{Wal01}. Here were discuss the broadband photometry. 

The F606W and F450W images were dithered using the standard 0.5'' diagonal
pattern. We coaligned and median-filtered the images to
remove cosmic rays and, in the dithered images, the warm pixels.
The target was clearly visible in all images.

We determined the net counts using aperture photometry. The background
is determined in an annulus of selectable width surrounding the target,
with the inner edge of the background annulus offset from the extraction
radius by a selectable amount. The mean background level is determined
iteratively by rejecting all pixels more than 3$\sigma$ from the median.
The net extracted counts are highly sensitive to small fluctuations in
the mean background, which depends on the location of the background
annulus. In the F606W filter, the substantial background causes the
formal S/N of the extracted counts to fall below unity for extraction
radii of 11-12 pixels. The highest significance of the measured flux
occurs for a 2 pixel (0.09~arcsec) extraction radius, which necessitates
a substantial aperture correction. 
We determined the aperture correction empirically using nearby bright stars.
We actually measured the source flux within a 3 pixel (0.14~arcsec) because
the aperture correction is more subject to errors arising from the exact
location of the star relative to the central pixel when using the 2 pixel
aperture.
We converted the count rates to fluxes
using the conversion factor given in
the PHOTFLAM keyword. We applied the
empirical aperture correction to the standard 11 pixel radius, and then
added the 0.1~magnitude correction to an infinite aperture. All
geometric corrections are negligible. We corrected for the charge-transfer
efficiency following \citep{Dol00}.
We apply decontamination corrections (1--3\%) to the F300W and F170W magnitudes.
We list these measured fluxes in Table~\ref{tbl-hst}, along with the formal
significance (in $\sigma$) for the detection. The uncertainty in the
F606W and F450W fluxes is dominated by the
uncertainty in the empirical aperture correction.


\subsubsection{Other Optical Observations}

We observed the target with the {\it NTT}
at the European Southern Observatory (ESO) on the night of
1997 August 9/10.
The night was photometric, but the seeing conditions (varying between
1.1 and 1.4~arcsec) required us to use
the ESO Multi-Mode Instrument (EMMI) instead of the
SUperb Seeing Imager (SUSI). We used the EMMI red CCD \#~36,
with the V-band filter ESO \#~606 and with the R-band filter
ESO \#~608. Throughout the night
we observed Landolt standard star fields for photometric calibration.

We took several images to reduce the risk of cosmic ray contamination
and placed the expected target position onto slightly different areas
on the chip in each exposure to avoid problems with bad pixels.  After
bias and flat field correction, we added the images using standard
MIDAS procedures to construct the final V-band image with a total
exposure time of 150 minutes.  This image is shown in
\cite{Neu98}. The target is clearly detected with a S/N of 18 inside
the {\it ROSAT} error circle at a position consistent with the {\it HST}
position.

We measured the $V$ magnitude to be $25.70\pm0.22$ 
(using the MIDAS command
magnitude/circle), and $V \simeq 25.72$ (using the MIDAS Romafot package).
This is consistent with the {\it HST} F606W flux.

We also obtained images in the Kron-Cousins $R$-band, totaling
two hours of exposure, towards the end of the night with air masses between
1.20 and 1.28, and seeing between 1.5 and 1.9 arcsec. We did not detect 
the target, with an upper limit of $R \simeq$24.5 mag.

\subsubsection{UV Spectrophotometry}

We obtained a far UV (1150-1720 \AA ) low dispersion (G140L) spectrum using the
{\it HST} Space Telescope Imaging Spectrograph (STIS; 
Woodgate et al.\/ 1998) on 23 and 26 October 2000.
We placed the target in the F52X0.5 slit using
a blind offset from star J (Walter et al.\/ 1996). 
The total integration time
is 26900 seconds (10 spacecraft orbits). The position angle of the slit was
359.4$^\circ$ and 358.5$^\circ$ on the two days.

We coadded the individual 2-dimensional spectra to 
produce a net spectrum.
There was no evidence for any spatial extent in this coadded image.
We extracted the source flux over 7 spatial pixels ($\pm$0.085~arcsec).
We extracted the background over 10 spatial pixels (0.244~arcsec), both above
and below the source, with an
offset of 2 spatial pixels between the extraction regions. 
We subtracted the background spectrum, and then smoothed the net spectrum.
The spectral shape is consistent with a blackbody. There are no statistically
significant spectral features. The largest spectral feature, at 1595\AA, has
an equivalent width of about 7$\pm$3~\AA.
We discuss this spectrum further in \S 6.

\section{Timing}

\cite{WWN96} reported the failure to detect a rotation period in the
{\it ROSAT} PSPC observation of RX~J185635-3754. The detection of a period
would be significant, in that it would require that the surface flux
distribution be non-uniform, or, if the surface is axi-symmetric about
some axis, that the axis be inclined with respect to the rotation axis.
In addition, the period would let one make inferences about the 
rotational history of the system, and search for a secular spindown. On
the other hand, the failure to detect rotational modulation could place
constraints on the magnitude of surface inhomogeneities. We report
here on a search for the pulsation period in the three {\it ROSAT}
data sets.

We searched each of the three {\it ROSAT} observations
(Table~\ref{tbl-rosobs}) for evidence of periodic modulations. We used
the FTOOLS tasks {\it abc} and {\it bct}  to generate and apply the
barycentric corrections to the data event lists.  We then searched the
data both by FFT and period-folding techniques. We verified the
techniques by recovering the correct periods and modulation profiles for
the Crab pulsar and the 8.4 s period X-ray pulsator RX~J0720-31
\citep{Hab97}.

The {\it ROSAT} data are not continuous,
but consist of a number of short segments.
We edited out the first 30 s of each segment, to mitigate against
the occasional effects of delayed high voltage turn-on, and ignored any
data segment shorter than 300 s.

\subsection{Power-Spectrum Analysis}

We generated an FFT of the data in each continuous time interval  and
then added the individual power spectra to generate a net power spectrum
for the complete data set. In this way we did not have to deal with the
data gaps. By ignoring the shorter intervals, we reduce the total
exposure time by about 10\%, but with little loss in sensitivity.
This reduced the net exposure time for RORs 400612 and 400864
by 15\% and 7\%, respectively; the PSPC observation was not affected.
 
We searched for periods between 0.1 and 200 s. No obvious signal stands
out in the power spectra of the three {\it ROSAT} observations.  (There is
strong power in the PSPC power spectrum at 80 seconds, but this is most
likely attributable to the spacecraft wobble, as the source is near the
inner supporting ring.  No such power is seen in the HRI data.)  If
there is a true rotational signal present, it will be present in all
three data sets, but  there are no coincidences amongst the 20 strongest
peaks in the three data sets. To determine the significance of the
strongest peaks in the power spectra, we generated 1000 (HRI) and 4000
(PSPC) simulated data sets with the mean count rate of the target. The
photon arrival times were uniformly distributed within the good
observation times. We added Poisson noise to the photon arrival times,
and generated power spectra.  The strongest peak in the PSPC power
spectrum has a 69\% likelihood of arising by chance. Similarly, the
strongest peak in the HRI power spectrum has a 90\% likelihood of
arising by chance.

\subsection{Period-Folding Analysis}

In a second search for a rotational period, we folded the data on trial
periods ranging from 0.1 to 12 s. The sampling periods were chosen such
that no period would be smeared by more than 10\% over the up-to-6-day
length of the observations. As we are searching for a smooth
(sinusoidal) modulation, and not a sharp pulse, this is acceptable. The
data were folded into 20 phase bins at each trial, and then tested
against the null hypothesis using a $\chi^2$~test. For each of the three
data sets, only a single period was found to yield a probability less
than that at which one would expect a single false alarm, but as the
three periods do not agree, we conclude that there is no detected
period.

\subsection{Limit to the Modulation Amplitude}

In order to determine our sensitivity to low amplitude periodicities, we
 generated simulated data sets with a sinusoidal modulation of known
amplitude. We varied the period of the modulation between 0.1 and 20 s,
and the amplitude from 4 to 15\%. We applied the power-series and
period-folding analyses to each simulated data set. We considered the
simulation a success if the input period was recovered to within the
uncertainty of the measured period, and a failure otherwise.

The observations recover the correct period over 50\% of the time for
amplitudes greater than about 6\% at periods shorter than 5 s, and
recover periods with amplitudes greater than 5\% for periods between 5
and 20 s. 
Thus we place an upper limit of 6\% on the amplitude of any rotational
   modulation in the soft X-rays.
This is about half the amplitude of the pulsation seen in 
RX~J0720-31 \citep{Hab97}.

\section{ATMOSPHERIC MODELING}

Neutron star atmosphere models for low magnetic fields ($< 10^{11}$ G)
were first developed by \cite{Rom87}, and followed later by others
\citep{Mil92,RR96,ZPS96}.  The models cover different compositions,
such as pure hydrogen, helium, carbon, nitrogen and iron, as well as a
solar mixture.  The strong 10$^{12-13}$G magnetic fields of pulsars
suggests that most, if not all, neutron stars should have similarly strong
fields. However, detailed atmosphere models with strong magnetic
fields are only available for hydrogen \citep{Zav95}, mainly because
reliable opacities and EOS have not yet been developed for heavier
elements.  For heavy-element dominated atmospheres, only approximate
treatments of magnetic Fe atmospheres exist \citep{RRM97}, and the
results show that the spectra are globally much closer to a blackbody
than for light element atmospheres.  We focus on non-magnetized models in
this paper, which affords a comparison with earlier work, and
also provides a benchmark for future calculations with magnetized
atmospheres.

We selected a small set of representative chemical compositions: pure H,
pure He, pure Fe, and a mixture of heavy elements labelled Si-ash. 
Table \ref{tab1} identifies the specific composition of the Si-ash case,
which was chosen to mimic the composition at the end of silicon burning
\citep{Arn96} and which might be typical of the matter initially
accreted onto a newly formed neutron star.  

We note that only a tiny amount of accreted matter from the
interstellar medium, $\sim 3 \times 10^{-4}$ g~cm$^{-2}$ of H,
corresponding to an accretion rate of $\dot M= 10^{-30}$ M$_\odot$~yr$^{-1}$
for a million years, is needed to render the atmosphere optically
thick to H at an energy of 0.25 keV.  Gravitational settling
ensures that heavy elements
settle out of the atmosphere and will not contribute to the emergent flux
in this event.  Nevertheless, it is not certain that all neutron stars
can accrete sufficient amounts of H for this to occur.  For example,
magnetized, rotating neutron stars will be in the ejector or propeller
phase \citep{Col98}, and are not expected to accrete.  Therefore, it is a
distinct possibility that some neutron star atmospheres are dominated
by heavy elements.

\subsection{Model Computation}

The atmosphere models have been calculated using standard techniques for
the construction of radiative, local thermodynamic equilibrium,
plane-parallel atmospheres \citep{Rom87,Mil92,RR96}.  The opacities and
EOS were obtained from the Los Alamos opacity project ({\tt
http://www.t4.lanl.gov}), prepared in tabular form.  A reference optical
depth (chosen to be a Rosseland mean) grid of 120 points is chosen for
the range from $10^{-8}$ to $10^2$, and photon frequencies are gridded
logarithmically in 190 levels from approximately 1 eV to 10 keV. 
Initially, hydrostatic equilibrium is imposed for a temperature profile
derived from the gray opacity approximation.  The flux is then computed
at each level of the atmosphere, and a variation of the Lucy-Uns\"old
procedure is used to calculate the temperature corrections needed to
satisfy flux constancy.  This variant utilizes the flux and Eddington
factors evaluated by numerical integration of the current profiles,
instead of using constant values of $1/2$ and $1/3$ as in the original
method.  This modification speeds convergence and accuracy.  With the
temperature corrections, hydrostatic equilibrium is reestablished, new
fluxes are evaluated, and the procedure is repeated until flux constancy
is reached to within 0.5\% throughout the atmosphere.

The monochromatic opacity used in the calculations is the sum of the
absorption ($\alpha_\nu$) and scattering ($\sigma_\nu$) opacity.
The frequency dependent source function can be written as
\begin{equation}
S_\nu = \frac{\sigma_\nu J_\nu + \alpha_\nu B_\nu}{\alpha_\nu + \sigma_\nu}
= (1-f_s) B_\nu + f_s J_\nu
\end{equation}
where $J_\nu$ is the mean intensity, $B_\nu$ the Planck function at
the local temperature, and $f_S={\sigma_\nu}/{(\alpha_\nu + \sigma_\nu)}$ is
the ratio of the scattering opacity to the total opacity.
This standard form of the source function has been used by
other authors (e.g. Zavlin et al.\/ 1996), and implies the isotropic
scattering approximation (no angle dependence in the opacities) which
has been shown to be accurate even in the case of scattering dominated
atmospheres (e.g. Mihalas 1978). Notice, however, that at the temperatures
of interest to our work the true absorption opacities are several orders
of magnitude larger than the scattering opacity ($f_s \ll 1$), especially for
heavy
element atmospheres. Thus the use of an even simpler source $S_\nu=B_\nu$,
as in Rajagopal \& Romani (1996), will suffice to obtain accurate spectra.
This situation can be different for low temperature ($10^5$K), light element (H)
atmospheres, where scattering processes contribute significantly and
corrections from anisotropic scattering might be significant.

Note that the redshift, which is significant in neutron star
atmospheres, does not enter into the calculation of the emergent flux.
The flux observed at the Earth, however, must be corrected for the
surface redshift of a neutron star.

\subsection{Comparison with Previous Work}

In Figure \ref{comp} we show the spectral fluxes of the emergent
radiation for selected effective temperatures and chemical compositions.
The BB fluxes are also shown for comparison.  In the left
panel we compare the BB flux (dotted lines) with that of pure Fe
atmospheres (solid lines), for the indicated values of $\log T_{eff}$. 
The range chosen, $5.25~<$~log $T_{eff}~<$~6.25,
allows a comparison with Figure 2 in
\cite{RR96}.  The agreement between our results with those of
\cite{RR96} is good, despite the fact that \cite{RR96} used opacities
and EOSs from the OPAL project \citep{IR96,RSI96}.  In the right panel,
we compare BB (dotted lines) and pure Fe (solid lines) fluxes with those
for pure H (dash-dotted lines) and Helium (dashed lines).  
For the right panel, we chose
specific values of $T_{eff}$ to allow a comparison with Figure 5 in
\cite{ZPS96}.
Although the agreement between H and He models in both
works is excellent, there are deviations apparent in the high energy
tail of the Fe models, especially for high temperatures.  These
differences do not appear to be explained by our use of Los Alamos
opacities, since both Zavlin et al.\/ (1996) and
\cite{RR96} employed OPAL
opacities.  Pavlov (private communication)
recently informed us that Zavlin et al.\/ (1996)
contained an error in the implementation of Fe opacity tables that might
explain the observed differences, and that corrected Fe atmosphere
models showed excellent agreement with those of \cite{RR96}, despite the
use of different algorithms for the atmospheres.

\subsection{The Effect of Gravity}

Aside from the effective temperature and the composition, the local
gravitational acceleration
 \be g = \frac{GM}{R^2 \sqrt{1-2GM/Rc^2}}
\ee potentially affects the emergent spectra.  This parameter is nearly
constant throughout the atmosphere because both its mass
and thickness are negligible compared to
the star's total mass and radius.  That the thickness of the
atmosphere is negligible compared to the stellar radius also justifies
treating the atmosphere as being plane-parallel, and this approximation
is generally even more accurate for neutron stars than for normal stars.

Figure \ref{fig_grav} shows the variations with $g$ of the emergent
(unredshifted) spectra of an Fe atmosphere with $\log T_{eff} = 5.6$.
The solid line is the emergent spectra of a canonical neutron star 
($M=1.4{\rm~M}_\odot$ and $R=10$ km, which corresponds to
$g_{14}=2.43$, where $g_{14}$ 
is the gravitational acceleration in units of
10$^{14}$ cm~s$^{-2}$). The emergent spectra
from two other configurations, one more compact ($R=$ 8 km,
$g_{14}=4.2$) and the other less compact ($R=$ 12 km, $g_{14}=1.6$),
are also shown. For reference, the BB spectrum, which does not depend
upon $g$, is also shown.  The effects of gravity on the emergent
spectra, in this energy range, are obviously small.  The chief differences
are in the vicinity of high-energy spectral lines and in the
high-energy tail, but these are too small, given the limited
resolution of the existent X-ray spectra, to affect parameter
constraints.

For the range of $g$ above, the corresponding gravitational
redshifts $z$ are $0.235<z<0.348$, where $z$ is defined through \be (1 +
z)^{-1} = \sqrt{1-{2GM\over Rc^2}}. \ee 
The redshift produces measurable
changes in the observed fluxes, as shown below.  Thus, although
separate constraints on $g$ and $z$ are possible in principle, only
$z$ is meaningfully constrained by atmospheric modelling at this time.
In the following sections detailing the results of spectral fitting,
the gravity is not treated as a free parameter.

\section{FITTING TO UNIFORM TEMPERATURE MODELS}

In this section, we focus on the estimation of parameters from spectral
fitting, assuming that the effective temperature is uniform across the
stellar surface.  However, it should be noted that a neutron star with a
moderate and non-uniform surface magnetic field, for example, possibly
has a non-uniform surface temperature as well, owing to the changing
conductivity of the surface layers.  In addition, hot spots on the
surface due to accretion or other phenonema may exist. The consequences
of allowing the temperature to vary across the surface of the star is
explored in \S 6.

The relevant parameters for spectral fitting
include the atmospheric composition, the
temperature observed at the Earth $T_\infty= T_{eff}/(1+z)$, the
redshift $z$, the interstellar medium column density $n_H$, and the 
angular diameter $R_\infty/D$.
The normalization factor $(R_\infty/D)^2$ is
the solid angle subtended by the star's surface visible at a distance $D$. 
The definition of $R_\infty=R/\sqrt{1-2GM/Rc^2}$ arises from the
blackbody relations, $F=4\pi$(R/D)$^2 T_{eff}^4$ and
$F_\infty=4\pi(R_\infty/D)^2 T_\infty^4$,
where $F_\infty$ is the flux observed at the Earth.
Note that the actual radius of the star $R$ must be less than $R_\infty$, and
that a measured value of $R_\infty$ also implies an upper limit to the
mass, $M<(c^2/G)R_\infty/\sqrt{27} \approx 0.13 M_\odot (R_\infty$/km).

The presence of an atmosphere can alter the thermal emission from a
neutron star substantially from a pure Planck spectrum.  For a pure
Planck spectrum, the redshift contributes only to an overall scale
factor, so that it no longer serves as a parameter and no information
concerning it can be obtained from fitting. For a realistic
atmosphere, however, the presence of spectral features, such as the
high energy cut-off observed in heavy-element dominated atmospheres in
\S 4, makes the redshift a measurable parameter.  In addition,
although we do not consider them in any detail in this paper, strong magnetic
fields, if present, could have an appreciable effect on the results of
atmospheric fitting (Zavlin et al.\/ 1995, Rajagopal et al.\/ 1997).

\subsection{Fitting the ROSAT PSPC Data}

We first examine the X-ray data from {\it ROSAT} since this is where the bulk
of the flux is found. Walter et al.\/ (1996)
discuss a blackbody fit to the PHA
spectrum. We have re-extracted the spectrum using FTOOLS and fit the data using
XSPEC V11.0.
We tabulated the neutron star
atmospheric model spectra as external additive tables for input into
XSPEC, according to the standard OGIP directives.
We fit only PHA channels 11-100 (0.11 to 1.0~keV). There is negligible flux
at higher energies, and the calibration in the lowest channels is questionable.

\subsubsection{Fits at the Nominal Redshift}

To facilitate comparison with previous work, we first fixed the 
mass $M$=1.4~M$_\odot$ and radius $R$=10~km, which gives 
$z$=0.305 and $R_\infty$=13.05 km.  The angular diameter
$R_\infty/D$, computed from the normalization of the model fits,
implies a value for $D$, which can be compared to
the recently measured parallactic value 
$61^{+9}_{-8}$ pc \citep{Wal01}.  Discrepancies between this and the predicted
values indicate that the atmosphere model, or the assumed redshift, are
incorrect.  Our results for the BB, H, and He atmospheres
substantially agree with those previously published (Walter
et al.\/ 1996, Pavlov et al.\/ 1996).
Our Fe atmosphere fit also gives a considerably lower
temperature and higher column density than that of \cite{Pav96}.
The difference is probably explained by deviations in the hard X-ray
tails of the emergent spectra noted in the previous section.
In Table~\ref{tab2} we
summarize the optimum fits to the {\it ROSAT} PSPC data for various
assumed compositions.

We also computed models with the Si-ash composition (Table \ref{tab1}), 
the results of which are intermediate between those of Fe and BB 
models,  probably because the presence of many elements and more absorption
lines causes a larger
energy redistribution than in a pure Fe atmosphere. The results from Table
\ref{tab2} indicate that the Si-ash and Fe spectra are reasonably
consistent with the measured distance.  

Note that none of the fits are formally acceptable. The best fit, that of the
blackbody spectrum, is rejected at the 99.5\% confidence level. The best fits 
for the various models are not statistically distinguishable: no single one
is preferable.
Figure \ref{fitFe} shows the best fit Fe atmosphere (which is formally the
worst of our fits). The residuals below 0.2~keV are present in all the fits,
and may be attributable to uncertainties in the PSPC response at low energies
(the PSPC response is not measured shortward of the Boron K$\alpha$ edge at
0.188~keV, but has been extrapolated based on the known characteristics
of the detector).
The high point at 0.3~keV is a statistical fluctuation.
The differences between the fits lie primarily between 0.7 and 0.9~keV,
where the Fe and Si-ash models underpredict the flux.

The predicted Si-ash and Fe spectra include a number of spectral lines.
These lines are not directly visible with
{\it ROSAT}, owing to the poor energy resolution ($\Delta E/E \approx 60\%$)
of the PSPC detector; however, they alter the spectral shape sufficiently
that they may constrain the gravitational redshift.
These absorption lines should 
be detectable in {\it CHANDRA} and {\it XMM-Newton} grating spectra.

\subsubsection{The Effect of Varying the Redshift}

The gravitational redshift $z$ is a function of $M/R$, and so to
fix $z=0.305$ is to presuppose this ratio (0.139~M$_\odot$/km).
While the BB results are unaffected by variations in
redshift, the strong absorption feature near 0.6~keV in the heavy element 
spectra, which acts as a
high energy cutoff to the emergent spectra, can in principle be constrained by
the X-ray spectra.

We used XSPEC to fit a series of redshifted models to the PSPC data.
We found that the best fit
to the data using the pure hydrogen model is not significantly
affected by the gravitational
redshift, probably because there are no discrete features in the emergent
spectrum. For both the Fe and Si-ash models we could reduce the value of
$\chi^2$ to that seen in the BB and H spectra by varying $z$, but the 
improvements
are not significant ($\Delta\chi^2\approx$20--30 for 86 degrees of freedom). 
The best fits with $z$ as a free parameter are presented in Table~\ref{tab2}. 

For both the Fe and Si-ash models, we find a strong correlation between
the best fit values of $n_H$ and $T_\infty$ (Figure~\ref{fz1}).
The general trend is that $n_H$ increases, and $T_\infty$ decreases, with
decreasing $z$
For a given $z$, the Si-ash
models give systematically larger temperatures (by about 6.5 eV) than
the Fe models.  

Quantitatively, the constraint between $n_H$ and $z$ appears to be
relatively insensitive to composition, and can be expressed as
\be n_{H,20} = (4.9\pm0.3) (1+z)^{-1}-(1.9\pm0.2)\,. \label{nhz} \ee
Separate correlations between $T_\infty$ and $n_H$ can be
extracted as well (Figure~\ref{fz2}).
The redshift $z$ decreases from right to left in this
figure.  The result for the BB model is also shown (diamond) for
reference.  The solid lines represent the quadratic fits
\be  T_\infty = \left\{\begin{array}{lc}
(157\pm2) - (96\pm3) n_{H,20} + (17\pm1) n_{H,20}^2{\rm~eV}&\qquad\qquad {\rm Fe} \\
(162\pm4) - (91\pm4) n_{H,20} + (15\pm1) n_{H,20}^2{\rm~eV}&\qquad\qquad {\rm Si-ash}
\end{array}\right\}\,.\label{ftnh} \ee
For a given redshift the 95\%
confidence contours are ellipsoids with a typical
vertical extent of about 5 eV above and below these curves.

The constraints in equations (\ref{nhz}) and (\ref{ftnh}) arise from the
competition between $T_\infty$ and $n_H$ in fitting the low-energy side,
and between $T_\infty$ and $z$ in fitting the high-energy side, of the
X-ray spectrum. A further constraint arises from the fact that the
spectrum can be approximated by a blackbody, so the total flux,
most of which is emitted in soft X-rays, is
proportional to $T_\infty^4(R_\infty/D)^2$.
We parameterize the effect of interstellar extinction
very simply as $\exp(-a~n_H)$, where $a$ is a constant, so that the observed
flux is proportional to $T_\infty^4(R_\infty/D)^2 \exp(-a~n_H)$. This works
because the strong interstellar extinction in the extreme ultraviolet
coincides with the peak of the thermal spectrum.
By so doing, we find the following
effective constraints for each composition:
\be T_\infty^4(R_\infty/D)^2 =\left\{\begin{array}{lc}
(6147\pm839)\exp[(1.46\pm0.07)n_{H,20}]~{\rm (eV)}^4 &\qquad {\rm Fe} \\
(3719\pm842)\exp[(1.58\pm0.10)n_{H,20}]~{\rm (eV)}^4 &\qquad {\rm Si-ash}
\end{array}\right\}\,\label{trnh} \ee
where $R_\infty$ is in km and $D$ is in pc.

The fits to the PSPC data provide interesting contraints on
the parameters $z$, n$_H$, $T_\infty$, and $R_\infty/D$. However,
the PSPC spectra alone can be used neither to decide the
atmospheric composition, nor to exclude any realistic values for
the gravitational redshift. For that we require the additional leverage
provided by the longer wavelength data.

\subsection{Fits to Multiwavelength Observations}

Figure \ref{comp} shows that the spectral energy distributions are quite
sensitive to both the composition and the temperature. In particular,
heavy element atmospheres have more opacity at short wavelengths than do
light element atmospheres, deviate less from a blackbody, and have a
smaller optical-to-X-ray flux for a given $T_{eff}$.
The X-ray band is sensitive to the exponential end of the Planck
distribution, as modified by interstellar extinction. 
Primarily because of the low spectral resolution of the PSPC, but also because
the low temperature ensures that the peak of the spectral energy distribution 
is highly absorbed, the extant X-ray data
are not useful for discriminating between the various models. However,
extrapolation of the optimum X-ray fits into the optical region
provides much greater leverage.
In this regime all the models can be approximated as the 
lightly reddened Rayleigh-Jeans tail
of a blackbody, with the observed intensity constraining
$T_\infty(R_\infty/D)^2$.

Here we incorporate into our analysis the {\it EUVE, NTT}, and
{\it HST} data summarized in \S 2.  
The dashed lines in
Figures \ref{figbb}--\ref{figH} show the extrapolations
of the best $z$=0.305 (Table~\ref{tab2}) {\it ROSAT}
PSPC fits to the full spectral energy distribution.
The {\it ROSAT}
data themselves are not shown because the response matrix redistributes
the energy and a model independent flux cannot be recovered. 
For consistency with the XSPEC calculations,
we use the \cite{MM83} cross sections for the EUVE extinction.
The {\it EUVE} data are generally consistent with the {\it ROSAT}
observations; thus they
are extremely useful in the energy range 0.1--0.18 keV in which the
{\it ROSAT} PSPC is not well calibrated. 

These figures show that for the best fits to the X-ray data,
the BB model underpredicts the optical flux by a
factor of 2.2$\pm$0.4, the Fe model overpredicts the optical
flux by a factor of 4\ud{.8}{.6},
the Si-ash model prediction is just about right (overprediction by
1.3\ud{.3}{.2}),
and the H and He (not shown) models 
overpredict the optical flux by a factor of 30. This suggests that
a single-temperature Si-ash atmosphere may be appropriate. Simultaneous
fitting of the long and short wavelength data will yield better constraints
on the phyical parameters.

\subsubsection{Fits at the Nominal Redshift}

We fit the multiwavelength data in the following manner:
We used XSPEC to generate a grid of fits in the $R_\infty/D$, $T_\infty$
plane at the nominal redshift, $z$=0.305. At each point
we fit the absorption column and generate a goodness of fit ($\chi^2$)
statistic for the PSPC data. 
At each point in the grid we then generate the appropriate model atmosphere and 
determine the goodness of fit statistic for the optical and UV data.
We weight each of the 4 data sets (PSPC,
{\it EUVE}, {\it HST}/STIS, and UV/optical photometry) equally.
At each point we sum the
reduced $\chi^2$ for each of the 4 data sets to determine a best fit region.
These $\chi^2$ grids are shown in Figures~\ref{fig-bbgrid}--\ref{fig-Hgrid}.
The {\it EUVE} and PSPC data define
a region that is most sensitive to the absorption column. The {\it EUVE}
data are
consistent with the PSPC data, but do not further constrain the fit.
The optical photometry and the STIS spectrum, lying on the Rayleigh-Jeans
tail of the spectral energy distribution, define a region of approximately
constant $R_\infty/D$.
The best fits generally occur near the intersections
of these two regions.
The resulting constraints are given in Table~\ref{tbl-mwf}.
These best overall fits are shown as the solid lines in Figures
\ref{figbb}--\ref{figH}.
We quantify the goodness of the joint parameters by multiplying the 
likelihoods of the most contraining data, the PSPC and optical data. These
are given in Table~\ref{tbl-mwf}.
We determine the luminosities (Table~\ref{tbl-mwf})
by integrating the overall best fit models, assuming a 61~pc distance.

Because the optical and UV fluxes define the Rayleigh-Jeans tail of a hot 
blackbody, we observe the quantity $T_\infty(R_\infty/D)^2\phi$, where
$\phi$ is the transmission though the interstellar medium. We use the mean
flux in the STIS spectrum and the mean flux in each of the 4 WFPC2 filters.
Each data point independently fixes a value for $T_\infty(R_\infty/D)^2\phi$
on the assumption that the observed flux is proportional to $\lambda^4$.
Minimizing the scatter in $T_\infty(R_\infty/D)^2\phi$ with respect to n$_H$
(using the \cite{Sea79} extinction curve)
yields n$_H$=1.6$\pm$0.2~$\times$10$^{20}$ cm$^{-2}$ and
$T_\infty(R_\infty/D)^2$ = 0.60$\pm$0.01. We also fit the data to determine the
following constraints:
\be T_\infty(R_\infty/D)^2 = \left\{\begin{array}{ll}
   0.59\ud{.03}{.01} & \qquad\qquad {\rm BB} \nonumber \\
   0.80\ud{.09}{.05} & \qquad\qquad {\rm Fe} \nonumber \\
   0.74\pm0.04       & \qquad\qquad {\rm Si-ash}
\end{array}\right\}\,\label{fopt} \ee
\noindent where $T_\infty$ is in eV, $R_\infty$ is in km and $D$ is in pc. The
BB result agrees with the analytic result.  That the Fe
and Si-ash results are larger than the blackbody result is
a consequence of different best-fit values of n$_H$ (greater extinction requires
a larger $T_\infty(R_\infty/D)^2$ to produce the same observed flux), with 
perhaps some contribution from small deviations from a pure Rayleigh-Jeans tail.
These constraints, together with those in equations (\ref{nhz}--\ref{trnh}),
in principle allow a unique determination of the four fitting parameters for
each model.

A non-magnetic
Si-ash model provides an acceptable fit in all wavelength bands.
This model gives $R_\infty/D$=0.13~km~pc$^{-1}$, or $R_\infty$=7.8$\pm$1.2~km
for the 61~pc distance.
The likelihood that the optical and X-ray parameters are identical is 53\%.

The best multiwavelength BB model yields a poor fit to the PSPC data.
Figure~\ref{fig_euve} shows that the best
multiwavelength fit provides a better fit to the {\it EUVE} data (both fits are
acceptable at the 1$\sigma$~level). This may be attributable to
uncertainties in the {\it ROSAT} PSPC calibration at low energies. Note that
the {\it EUVE} detection corresponds to PSPC channels 11-16, which seem to
show systematic deviations in Figure~\ref{fitFe}.
The best fit to the
X-rays alone underpredicts the optical flux, but this
can be made up with emission from a component too cool to
contribute significantly to the X-ray emission.
The likelihood that the optical and X-ray parameters are identical is 0.03\%.
A multi-temperature blackbody model can yield an acceptable fit (see \S6). 

The best non-magnetic Fe model provides an unacceptable fit in the X-ray
band, but its parameters are similar to the Si-ash case. Unlike the case
of the BB, the best X-ray fit overpredicts the optical flux. An overprediction
cannot be lessened by adding additional cooler regions.
Additional opacity, as in the Si-ash atmosphere, brings the star closer
to the blackbody limit and would reduce the optical overprediction.
As we discuss in \S\ref{sec-mwz}, the overpredicted optical flux can
be lessened by an increase in $z$.

The uniform temperature non-magnetic hydrogen model can be excluded
by these data (Figures~\ref{figH} and \ref{fig-Hgrid}). 

\subsubsection{The Effect of Varying the Redshift\label{sec-mwz}}

The BB models are not sensitive to the redshift.
The hydrogen model is sensitive only in that 
the observed wavelength of H~I Lyman~$\alpha$ depends on $z$. The model
predicts that the line is
in absorption with an equivalent width of about 25\AA.
We see no significant absorption lines in the far-UV, with a limiting
equivalent width of about 10\AA, excluding H-dominated model atmospheres
in the range $0.07 < z < 0.38$. The He atmosphere similarly exhibits no
significant redshift dependence.

The heavy element atmospheres are affected by the redshift.
We discussed the effects of the redshift on the X-ray spectral fits earlier.
At long wavelengths the redshift dependence appears because the
flux is proportional to $T_\infty (R_\infty/D)^2$
(with a small correction
for reddening), and $T_\infty$ is a function of $z$. Higher redshifts, and
larger values for $T_\infty$, require smaller values for $R_\infty/D$.
In both the X-ray and optical regimes we
obtain relations between $R_\infty/D$ and $z$, as shown in Figure~\ref{fz3}.
The intersections of these regions constrain $z$ and $R_\infty/D$.
For the Si-ash model, we find that 0.124~$<~R_\infty/D~<$~0.143 and
0.305~$<~z~<$~0.344. For a pure Fe atmosphere the constraints are
0.124~$<~R_\infty/D~<$~0.149 and 0.346~$<~z~<$~0.405. 
Note that the values of $R_\infty/D$ are nearly identical.
These best fits for the Si-ash and Fe atmospheres, summarized in
Table~\ref{tbl-params}, are shown as the dotted lines in
Figures \ref{figFe} and \ref{figSi}. 

The uniform temperature non-magnetic hydrogen model cannot be saved by varying
$z$ because $T_\infty$ and $R_\infty$ do not have any significant redshift
dependence.

\subsection{The Best Fit Parameters}

Having parameterized the constraints from the X-ray fits
(\ref{nhz}--{\ref{trnh}), together
with the optical constraint in equation (\ref{fopt}), in principle
we can uniquely determine the four fitting parameters. This provides 
a consistency check on the parameters derived from the multiwavelength
$\chi^2$ minimization (Table~\ref{tbl-mwf}).

The best parameters for the heavy element atmospheres are summarized as:
\be
\begin{array}{l}
\left\{ \begin{array}{ccc}
R_\infty=8.2\pm1.3 {\rm ~km} & R=6.0\pm1.5 {\rm ~km}  
       & M=0.95\ud{0.06}{0.16} M_\odot \\  
z=0.39\pm0.12  & n_{H,20}=1.7\pm0.1 & T_\infty=44\pm4~ {\rm eV}
\end{array}\right\}
\quad{\rm Fe}\\

\left\{ \begin{array}{ccc}
R_\infty=7.8\pm{1.3} {\rm ~km} & R=6.0\pm1.4 {\rm ~km} 
       & M=0.84\pm0.07 M_\odot\\
z=0.31\pm0.12  & n_{H,20}=1.8\pm0.2 & T_\infty=45\pm6~
    {\rm eV}
\end{array}\right\}
\quad{\rm Si-ash}
\end{array} 
\ee
The quantities $R_\infty, R$, and $M$ all scale linearly with $D$.

We compare these analytic values to the fits to the multiwavelength
data in Table~\ref{tbl-params} for the two
heavy element atmosphere models. The uncertainties on the analytic fits
exceed those determined from the data directly, primarily because the
uncertainty in the direct measurement of the temperature is significantly less
than the scatter in the quadratic fit to $T_\infty$ as a function of n$_H$.
Within their uncertainties the derived values of the parameters are in
excellent agreement.

\subsection{Comparison to Theoretical Expectations}

In Figure \ref{mr1}, the inferred masses and radii for uniform
temperature models are compared with mass-radius curves for a wide
variety of equations of state.  The EOSs in this figure are labelled
following the convention of \cite{LP2000}, and include cases with
baryonic compositions (MS0, MS1, PAL6, GM3, AP4) together with one
with a kaon-condensed core (GS1), all indicated by solid curves.  In
addition, two cases of self-bound stars composed of pure quark matter
(SQM1, SQM3) are shown by dashed curves.  The diagonal dashed line
labelled ``causality'' is the approximate boundary $R>3.04GM/c^2$
imposed by the requirement that the EOS never violates causality, and
lines of constant $R_\infty$ are indicated by the dotted curves.
 
The cases MS0 and MS1 are representative examples of nucleonic
field-theoretic EOSs, and AP4 is a state-of-the-art non-relativistic
potential model EOS.  The cases GM3, GS1 and PAL6 were chosen because
of the softening they display above nuclear density: this is caused by
hyperons and a kaon condensate for the field-theoretical EOSs GM3 and
GS1, respectively, and by an extremely small incompressibility
parameter for the schematic potential EOS PAL6.  Note that all these
cases give values of $R\gtrsim10$~km for 1.4--1.5 
M$_\odot$, except in the case of extreme softening induced by a phase
transition ({\it viz.} GS1).  It is not inconceivable that even
more compact stars could be obtained, but simultaneously satisfying
the 1.44 M$_\odot$ 
mass constraint (of PSR~1913+16) becomes very difficult.
 
The EOSs SQM1 and SQM3 illustrate the different behavior, relative to
normal neutron stars, of self-bound configurations.  Self-bound quark
stars are subject to two constraints: first, that strange quark matter
is the true ground state of matter (the so-called Witten's conjecture;
Witten 1984), and second, that the maximum mass be larger than that
of the most massive, accurately measured, neutron star (PSR 1913+16).
The case SQM1 represents the most compact self-bound quark EOS that
obeys these constraints.  More compact configurations can be obtained,
but only by violating these constraints and thus cannot be considered
realistic in that connections to identifiable physics have not been
established \citep{PBP, LPMY}.
 
The crosses in Figure \ref{mr1}, one for each of three assumed
distances 51 pc, 61 pc and 71 pc, are the centroids of the allowed
(hashed) regions whose extents are determined by the uncertainties we
indicated in the four constraints (\ref{nhz})--(\ref{fopt}).  These
regions imply configurations that are too small, for their
masses, to be explained by a reasonable EOS, including those for
self-bound strange quark matter.  In addition, the masses are too small to 
fit theoretical expectations of neutron star masses from evolutionary 
considerations.  The
uncertainty in the distance is too small to affect these conclusions.   
These estimates of $M$ and $R$ are not
sensitive to the precise way in which an
optimum fit to the three sets of data ({\it ROSAT, EUVE}, and {\it HST}) is
determined, i.e., on how much statistical weight should be assigned to
each set of data.

We emphasize that a similar fit to all data is not
possible for non-magnetized atmospheres composed by light elements (H,
He).  The four magnitude overestimate of the optical flux for an H or
He atmosphere is too large to be reconciled without unacceptable
deviations from the X-ray data. 
\cite{Pav96} found that a magnetic field of 
$10^{12}$ G results in a decrease of 1.5 magnitudes ($V$ band), 
compared to the non-magnetic case, with $R_\infty/D \approx 1.1$~km~pc$^{-1}$. 
This correction is still much too small (by a factor of 20)
to reconcile with the observed 
optical flux. More recent models with higher magnetic fields,
appropriate for magnetars, seem to require hard power-law tails
\citep{HL01,Oz01}, and even a deficit of low energy photons, relative
to the best blackbody fit, neither of which are observed in this case
(although their models are for hotter sources).
\cite{ZTT00} model accreting magnetized atmospheres and can reproduce
the characteristic optical excess and lack of a hard tail, but at the
expense of a luminosity two orders of magnitude larger, and temperature
one order of magnitude, larger than we observe. It seems unlikely that
a magnetized hydrogen atmosphere can be reconciled with the observed 
spectral energy distribution.

However, X-ray spectra of neutron stars are often successfully fit with
hydrogen or magnetic hydrogen atmospheres.
In some cases hydrogen atmospheres are expected because the
neutron star is accreting in a binary system (e.g., Rutledge et al.\/ 2001).
In other cases it may be that
the X-ray spectrum alone provides insufficient leverage to distinguish between
competing models (as is the case here).

\section{NON-UNIFORM TEMPERATURE MODELS}

Given the small value of the radius obtained from models
in which the entire surface is assumed to be at the same temperature, we
assess here the qualitative changes expected
when non-homogeneous temperature distributions are considered. Our model
atmospheres were built under the assumption of isotropic surface emission
and low magnetic fields. The presence of a large magnetic field might cause
significant anisotropy in the energy transfer. It has been
shown that even without a magnetic field, the emerging spectrum will vary with
the viewing angle with respect to the atmosphere normal. For emission from
a small region of the star, these limb darkening effect causes substantial
variation (see e.g. Zavlin {\it et al.}, 1996). Thus the validity of our
models is constrained to uniform or near-uniform temperature distributions.
The careful modelling in the case of temperature anisotropies is beyond the
scope of this paper. With this caveat in mind, we 
infer some qualitative results and indicate the expected trends by exploring
simplified two-temperature models.

We consider here a simple
two-component model in which a hot polar cap accounts for the
X-ray emission and a cooler equatorial region explains the optical
data.  We denote the hotter temperature component with the subscript
$H$ and the cooler component with the subscript $C$, respectively.
Then the fractional area covered by the hot component is
$\alpha=(R_{H\infty}/R_\infty)^2$ and
$R_\infty^2=R_{H\infty}^2+R_{C\infty}^2$. 

\subsection{Two-Component Blackbody Models}

A reduction in the fractional surface area
of the hot region results in an increase in the optical flux.  
As an example, we consider a blackbody model with the component
temperatures arbitrarily set to $T_{H\infty}=55.3$ eV, the best-fit
temperature for the uniform temperature blackbody model, and
$T_{C\infty}=20$ eV.  In Figure~\ref{twobb} we show the spectral energy
distribution for the one component model ($\alpha=1$; dashed curve) and
for a model with $\alpha = 0.2$ (solid curve),
which fully accounts for the optical emission.
This behavior is easy to understand.  The
X-ray flux from the hot component dominates that of the cool one.
But the optical flux from the hot component alone underestimates the
measured optical flux by a factor $f$, which in the present case is
$f=2.3$.  One then has
\be f-1={T_{C\infty}\over T_{H\infty}}
\left( {R_{C\infty}\over R_{H\infty}} \right)^2\,.   \ee
The definition of $\alpha$ leads to
\be \alpha=\left[ 1+(f-1){T_{H\infty}\over T_{C\infty}} \right]^{-1}
\simeq(1+3.6)^{-1} \simeq0.2\,, \label{alp1}  \ee
for the example described above.  
The fractional
contribution of the  cool component to the total flux is
\be \left({T_{C\infty}\over T_{H\infty}}\right)^4
\left({R_{C\infty}\over R_{H\infty}}\right)^2=
(f-1)\left({T_{C\infty}\over T_{H\infty}}\right)^3\simeq 0.06\,.\ee 
Since the cooler component peaks at 
longer wavelength, this is a strong upper limit to the fractional
flux in the X-ray band (In actuality, less than 0.2\% of the X-ray flux 
in this model arises from the cool component. In a model where
$T_{C\infty} = T_{H\infty}/2$ the cool component contributes 2\% of the
soft X-ray flux.).
This justifies using the same temperature for the hot component
as that of the uniform temperature model.
A small value of $\alpha$ results in a large increase
in the predicted emitting surface area. In the present case the value of
$R_\infty/D$, which is proportional to $1/\sqrt{\alpha}$, has more than
doubled.

Figure \ref{2bbc} shows the behavior of the inferred value of
$R_\infty$ as the temperature of the cool component is lowered even
more.  The values of $T_{H\infty}$ and $n_H$ are kept fixed and equal
to the best fit to the uniform temperature BB model, which ensures
reasonable agreement with {\it EUVE} and X-ray data. The inferred value of
$R_\infty$ can be enlarged to as much as $15$ km if $T_{C\infty}$ is
reduced to 5 eV.  The average visible surface area of the hot
component in that case is only 6.5 \% of the total surface of the
star. The results essentially follow the simple relation given by
equation (\ref{alp1}), which shows that as $T_{C\infty}\rightarrow0$, so
does $\alpha$.

In principle, it seems that one could make the star arbitrarily large
by a dramatic decrease in the temperature of the cool component.
However, the UV/optical spectrum arises in large part from the cooler
region, and this allows us to place a firm lower limit on its
temperature.  For a sufficiently cool temperature, the spectrum will
deviate from the Rayleigh-Jeans tail at the short wavelengths in the
near UV part of the spectrum. This is shown in Figure \ref{uvcons},
where we plot the far UV spectrum from the {\it HST}.

For a baseline, we subtract the best-fit, uniform-temperature BB model for
the X-ray spectrum (Table~\ref{tab2}),
and then fit the residual flux with a series of BB
spectra, normalized to the total flux between 1310\AA~ and 1650\AA~. 
The continuum slope of the far-UV spectrum cannot be reconciled with a
BB with $T_\infty<6.5$~eV at 90\% confidence.
If we further require that the BB curve pass through
both the UV and optical photometric points (Figure \ref{uvcons} inset),
then we can place a 1~$\sigma$ confidence lower limit of 15~eV
on the temperature of
any cool surface that dominates the optical flux (this is the dotted vertical
line in Figure~\ref{fig_alphaz}). 
The value of $R_\infty$ for a two-component BB model could thus be increased
relative to the uniform temperature BB model by, at most, the factor
$1/\sqrt{\alpha}$, which from equation (\ref{alp1}) for
$T_{H\infty}=55.3$ eV and $T_{C\infty}=15$ eV, is about 2.4.  This
corresponds to an upper limit to $R_\infty$ of about 10~km.

\subsection{Two-Component Atmospheric Models}

In \S\ref{sec-mwz}, we showed that uniform temperature models could fit
the data, with formal best fit values for $z$ of 0.34 (Si-ash) and 0.37 (Fe). 
Higher gravitational redshifts require a hotter and smaller surface, which
underpredicts the optical flux. This flux could be supplied by additional cool 
components which do not emit at X-ray wavelengths. Thus two-component heavy
element atmospheres are readily accomodated for higher redshift
surfaces.

We follow the formalism developed in the previous section, but introduce
the redshift dependence of $T_{H\infty}$. Figure~\ref{fig_alphaz} shows 
the permissable values of $\alpha$. To the left of the dotted
vertical line the models overpredict the optical flux. The region below
the dash-dot line is excluded by the slope of the optical flux (\S6.1).
The dashed line marks where $T_{C,\infty}$ is half $T_{H,\infty}$; we
exclude the region above this line since the shape of the X-ray
spectrum will be affected by such a hot component. The allowable
region lies between these bounds; in this region $\alpha$ increases with
increasing $z$. 

Using the parameterization of $R_\infty/D$ with $z$ for these models, one can
use these values for $\alpha$ to estimate the maximum $R_\infty/D$ to be
about 0.21 km~pc$^{-1}$ for both models (Figure~\ref{fig_maxrad}).
These represent only 60\% increases in the radius over the best 
single component radius. The maximum radius occurs at high $z$, about 0.6;
the increase in radius with $z$ occurs
because of the interplay between the increase in $\alpha$ and the decrease in
$R_\infty/D$ with increasing $z$. We conclude that one cannot increase the
radius arbitrarily, and that there is a clear upper limit on $R_\infty/D$
of about 0.21~km~pc$^{-1}$ for these heavy element atmospheres. It is
possible to approach the 10~km radius of the canonical neutron star
using non-magnetic heavy element atmospheres. Given $R_\infty/D$ as a 
function of $z$, the minimum mass for the neutron star occurs near the
minimum $z$, and the mass increases as $z$ is increased, exceeding the 
canonical value of 1.4~M$_\odot$ for $z \gtrsim 0.5$ (Figure~\ref{fig_az3}).

It is also clear that the light-element atmospheres cannot be made
consistent with the optical data by means of a non-uniform
temperature model.  
The primary reason is that the best fits to X-ray
data for these models overestimate the optical fluxes by a much
greater factor than do heavy-element models. The lack of a dependence
of $T_{H,\infty}$ on $z$ (\S\ref{sec-mwz})
means that one cannot remove the optical
excess by increasing $z$, and so there is never any optical deficit to be 
made up with emission from a cool component.

The four analytic constraints for a simple two-component
model are: \be 
T_{H\infty}(R_{H\infty}/D)^2+T_{C\infty}(R_{C\infty}/D)^2 =
\left\{\begin{array}{lc}
0.80\ud{0.09}{0.05}{\rm~eV} &\qquad {\rm Fe} \\ 
0.74\pm0.04{\rm~eV} &\qquad {\rm Si-ash}\end{array}\right\} \label{fopt2} \\
n_{H,20} = (4.9\pm0.3)(1+z)^{-1}-(1.9\pm0.2)\qquad\qquad\qquad\qquad \label{nhz2} \\
T_{H\infty} = \left\{\begin{array}{lc}
(157\pm2) - (96\pm3)n_{H,20} + (17\pm1)n_{H,20}^2{\rm~eV}&\qquad {\rm Fe} \\
(162\pm4) - (91\pm4)n_{H,20} + (15\pm1)n_{H,20}^2{\rm~eV}&\qquad {\rm Si-ash}
\end{array}\right\} \label{ftnh2} \\
T_{H\infty}^4(R_{H\infty}/D)^2+T_{C\infty}^4(R_{C\infty}/D)^2 =
\left\{\begin{array}{lc} (6147\pm839)\exp[(1.46\pm0.07)~n_{H,20}]{\rm (eV)}^4
&\quad {\rm Fe} \\
(3719\pm842)\exp[(1.58\pm0.10)~n_{H,20}]{\rm (eV)}^4 &\quad {\rm Si-ash}
\end{array}\right\} \label{trnh2} \ee 
In addition, we have the constraints
\be
R^2_{C\infty}>0 \label{frc}\\
T_{H\infty}/2 > T_{C\infty} >15~{\rm eV}\,. \label{fthc}
\ee
Note that constraint (\ref{frc}) is equivalent to setting a lower limit to
$z$.
Additionally, we limit possible values of $M$ and $R$ by causality.
As previously, radii are in km and distances in pc.

The allowable region in the mass-radius diagram from the two-temperature
model defined by equations \ref{fopt2}--\ref{fthc} is
shown in Figure \ref{mr3}.  The
shaded region indicates the allowed values of $M$ and $R$ at 90\% confidence,
for an assumed distance of 61~pc, including the uncertainty in the distance.
The results for $M$ and $R$ scale with $D$.  

As expected, the addition of the second component enlarges the allowed
regions for $M$ and $R$, but there exists a lower limit on the
compactness. We note that the largest radii occur for the most extreme
temperature variations and for the largest allowable distance. While
this simplified model shows the qualitative trends, only a more detailed, and
self-consistent, analysis of surface inhomogeneities can establish
realistic limits.

\section{DISCUSSION AND OUTLOOK}

We have calculated a series of neutron star model atmospheres for
different chemical compositions, which have been used as input models to
fit multiwavelength spectrophotometric observations of the nearby
compact object RX~J185635-3754.  We have investigated
the constraints that exist among relevant parameters which most
influence the predicted spectra.  The {\it ROSAT} X-ray data alone are insufficient
to adequately constrain these parameters, but the combination of X-ray and
UV/optical observations does allow significant constraints to be imposed.  
Our main conclusions are:

\begin{itemize}
\item A uniform temperature blackbody is excluded by the multiwavelength data.

\item Non-magnetized light-element atmospheres are
      excluded since they are incompatible with combined
      optical and X-ray observations.  Results for magnetic H-atmosphere
      models \citep{Pav96} indicate that fields of order $10^{12}$ G or less are
      unlikely to change this result. 

\item The simplest uniform temperature heavy-element atmospheric models
      indicate that $T_\infty\approx$ 45~eV, $R_\infty\approx$ 8~km,
      $M\approx$ 0.9~M$_\odot$, and $R\approx$ 6~km. This mass and radius are
      too small to be consistent with any equation of state in common use,
      including even that of self-bound strange quark matter.  This value
      for $R_\infty$ represents a lower limit in models with temperature 
      inhomogeneities.

\item For heavy-element compositions or a blackbody, the accumulated optical
      data yields- $T_\infty(R_\infty/D)^2=0.7\pm0.1$ eV because it falls on
      the Rayleigh-Jeans tail of the emission.  

\item That the optically flux does not significant deviate from a 
      Rayleigh-Jeans behavior implies that the temperature of the star is
      greater than 15 eV.  Combined with upper limit to
      the distance, this indicates that $R_\infty\le16.5$ km 
      ($M\le2.15$ M$_\odot$) at $1~\sigma$.

\item A simple two-component blackbody model provides an acceptable fit of
      the data for $R_\infty\le10$ km ($M\le1.3$ M$_\odot$).

\item Simple two-component heavy-element atmospheres can also provide 
      acceptable fits, provided $z$ exceeds 0.34 (Si-ash) or 0.37 (Fe). 
      In these cases, neutron star configurations up to $R\approx10$ km 
      can be allowed.  Further investigation of models including 
      inhomogeneous thermal emission are thus worth pursuing.

\item The uncertainty in the luminosity is less than those of any
      other neutron star for which thermal emission is believed to be seen.
      We find, for acceptable models, 
      $L=1.5\pm0.2\cdot10^{31}$ erg/s including the uncertainty in the 
      distance.  
      As shown in Figure \ref{lumrxj}, which summarizes the existing data
      concerning neutron star cooling, the location of RX~J185635-3754
      is consistent with both the
      spin-down luminosities and ages of the other objects shown. 

\end{itemize}


\subsection{Causes and Implications of Non-Uniform Temperature Distributions}

If the surface is a blackbody, or if radius is to approach the predictions
from realistic equations of state, then it appears that the surface temperature
must be non-uniform. Here we briefly discuss some possibilities for
generating these anisotropies,
and whether we should expect to detect them directly.

The magnetic field itself could generate a large temperature
gradient \citep{SY96} if its strength exceeded $10^{12}$ G. Note,
however, that under this assumption, the hot polar region contains
about 80\%-90\% of the surface area, while the analysis presented in \S 6
implies the opposite. Nevertheless, this scenario
might account for about a 20\% increase in the radius for a model
wherein the contribution of the hot component only slightly
underestimates the optical fluxes. This still leaves the estimated
masses and radii well below theoretical expectations.

Another possible source of temperature anisotropy is rotation.  \cite{MRL93}
investigated rotationally produced temperature
anisotropies, while the effect of rotation on the cooling of neutron
stars has been investigated by \cite{SW98}.  A neutron star with an
isothermal core, rotating with nearly its Keplerian frequency, might
have a polar temperature up to 30\% higher than the equatorial
temperature, due to the angular dependence of the surface gravity.
This effect also seems smaller than what is required.

A relevant question is whether or
not a neutron star with an inhomogeneous surface temperature,  spinning
with typical periods and with a magnetic field of the order  of
$10^{12}$ G, should be seen as an X-ray  pulsar.  As shown in \S 3, no
modulation has been detected at a level above 6\%. This is not, however,
a strong argument against a large surface temperature anisotropy since
a star as compact as we have inferred ($GM/Rc^2$=0.22--0.32) has
a maximum allowed pulse fraction of less than 10\%
\citep{Pag95,PS96,POD00}. This limit is even lower if the magnetic and
rotation angles are not perpendicular or if the hot spot occupies a
large fraction of the surface. The lack of modulation in the signal  is
due to general relativistic deflections which expose a large fraction of
the surface to an observer at infinity.  Any modulation detected in the
future by more sensitive observations would thus be extremely useful for
setting upper limits to the compactness of the object, as has been
claimed for other isolated neutron stars \citep{Wan99}.  A related
effect \citep{POD00} is that, for $z>0.25$, the
spectroscopically inferred polar-cap surface area is, at most, 10\%
different than its intrinsic area. Therefore our estimates of the
radius from the inferred surfaces are barely affected by the observer's
inclination or the opening angle of the polar cap.

\subsection{Can RX J185635-3754 Be Accreting?}

Accretion can be a source of heating, and is a mechanism which can
generate relatively large surface temperature variations. Earlier, we
dismissed Bondi-Hoyle accretion onto a non-magnetized star as an
unlikely source of the observed luminosity because of the large space
velocity (unless the local interstellar density is about
10$^4$~cm$^{-3}$).  However, as we also pointed out, it does not take
much accretion to make a hydrogen atmosphere.

An independent check of the accretion hypothesis is provided by the
{\it ASCA} data. \cite{Nel95} argue that a magnetized neutron star
accreting from the interstellar medium could emit 0.5 - 5\% of its
accretion luminosity in the form of a cyclotron line between 5 and 20
keV. The {\it ASCA} spectra, however, place an upper limit to the flux
between 1.5 and 12~keV of 0.5\% of the total X-ray flux. If the soft
X-ray flux were entirely due to accretion, we could exclude
magnetic field strengths between about 1 and
7$\times$10$^{12}$~G. Conversely, if the star is magnetized in this
range, then no more than about 10\% of the soft X-ray luminosity could
be due to accretion.  

If RX~J185635-3754 is a typical $10^6$ year old neutron star, then it is
most likely magnetized. At the $10^6$ year age 
suggested by its parallax and proper motion (Walter 2001), 
it is likely to be in the ejector phase \citep{Tre00}. If older,
then it is most likely in the propeller phase like most neutron stars
\citep{Col98}.  In either case, the large magnitude of the magnetic
fields implied by the ejector and propeller phases make it extremely
unlikely that it could accrete from the interstellar medium, and may explain
why the atmosphere is not dominated by hydrogen.

Another indication that this star is not significantly accreting, but
that it might be magnetized, is based on the deep VLT image released by
\cite{KK00}. This image shows a classic bow-shock nebula. The presence
of a bow-shock suggests that this is a magnetized neutron star with
a relativistic wind, as observed in the pulsars PSR 1957+20 \citep{KH88}
and PSR 2224+65 \citep{CRL93}.  Analysis of the broadband F606W images
of this region \citep{bows} show that the standoff distance between the
target and the apex of the nebula is about 1~arcsec (60 AU at a distance
of 60 pc).  Pressure balance between the interstellar medium and the
relativistic wind produces a standoff distance of 26
$\sqrt{B_{12}/n}/(P^2v_{100})$~AU, where $B_{12}$ is the magnetic field
strength in units of $10^{12}$ G, $n$ is the density of the interstellar
medium in cm$^{-3}$, $P$ is the rotation period of the neutron star in
seconds, and $v_{100}$ is the velocity of the neutron star relative to the
interstellar medium in hundreds of km~s$^{-1}$.  While other mechanisms
can cause a bow-shock geometry, their spatial scales are not correct. In the
absence of a relativistic wind, the ram pressure of the interstellar
medium on the neutron star's magnetic field leads to a standoff distance
of $0.02(B_{12}/(\sqrt{n}v_{100}))^{1/3}$~AU.  An ionization front has an
expected radius of $4000\sqrt{L_{31}/(nv_{100})}$ AU,
where $L_{31}$ is the luminosity of ionizing photons in units of
$10^{31}$~erg~s$^{-1}$, and there should be a partially ionized zone
extending out about twice as far.

\cite{bows} estimate from the VLT image that RX~J185635-3754 likely has
$B_{12}>0.4$~G, and $P>0.5$ s. These values are fairly typical for old
pulsars. In this case, RX~J185635-3754 might be a dead or misaligned
radio--pulsar. If so, RX~J185635-3754 has likely never accreted after
the initial fallback from the supernova, which could explain why the
surface is apparently dominated by heavy elements and not hydrogen. 
A typical pulsar magnetic field
may also not be large enough to substantially affect
the emergent spectra of heavy-element atmospheres \citep{RRM97}.
In fact, \citep{RRM97} show that the spectral energy distributions of
magnetized  ($10^{12}-10^{13}$ G) Fe atmospheres are closer to blackbodies
than are those of unmagnetized Fe atmospheres. In light of
the small differences between our Fe atmospheres and the BB model fits, it is
unlikely that the presence of magnetic fields 
of this magnitude will substantially affect our conclusions.  

\subsection{Epilogue}

The simplest interpretation consistent with the data presented here
is of uniform
thermal emission from a compact object with a heavy element atmosphere.
The object appears to be smaller than any canonical neutron star
or self-bound strange star and less massive than the current
supernova paradigm would allow.
Allowing surface temperature inhomogeneities still results in a
relatively compact object,
but one which no longer excludes all models. 
A two-component blackbody is acceptable, as are one or two-temperature
component heavy element atmospheric models.
The most robust results from the atmospheric modelling appear to be a
restriction of the star's redshift
($z>$0.3) and a surface composition devoid of
light elements.
The large inferred
compactness is not completely without precedent: for example, \cite{WKMO01}
have recently argued that very compact stars ($z\sim$0.58)
are required to explain the yield of $r$-process elements, if produced in a
neutrino-driven wind. 

The small inferred radius aside,
the star does not appear unusual. The bow-shock image
(\S7.2) suggests that this may be a common pulsar viewed from the
side. The luminosity (Figure~\ref{lumrxj}) appears normal for its age.

Data such as these open the exciting possibility that observations can
offer meaningful constraints on the EOS.  The uncertainties in the
values we have obtained for the mass and radius of RX~J185635-3754 are
mostly limited by the relatively poor spectral resolution of the
X-ray data.  Nevertheless, this object does afford a clear opportunity
to measure these properties accurately when higher resolution X-ray
spectra from the Chandra and XMM-Newton observatories become
available.  Most importantly, such spectra may reveal X-ray lines
which would allow the unambiguous determination of the star's
atmospheric composition and redshift. High S/N far-UV spectra in the
1500-2000\AA\ range can also be used to search for a H~I Ly~$\alpha$
line, which may be present even if the overall composition is
dominated by heavy elements. And if the spectra reveal no lines,
perhaps we should recall that the spectrum of a self-bound strange
quark matter star is likely to be a pure thermal spectrum.
RX~J185635-3754 has yet to reveal all its secrets.

\acknowledgements We thank Roger Romani, George Pavlov, and Lars Bildstein
for beneficial discussions and comments.
We acknowledge stimulating discussions with Ralph Wijers.
JAP gratefully acknowledges J.A. Miralles
for helpful discussions. 
We appreciate the careful review of an anonymous referee.
Kaisey Mandel contributed to the X-ray timing
analysis. This work was supported in part by the U.S.
Department of Energy under contract numbers DOE/DE-FG02-87ER-40317 and
DOE/DE-FG02-88ER-40388, the NASA grants NAG52863, NAG54862 ({\it ASCA}),
NAG57391 ({\it EUVE}), the LTSA grant NAG57978 and also grants GO 074080196A
and GO 081490197A from the Space Telescope Science Institute.
JML is appreciative of the support
of the J.S. Guggenheim Foundation for a Fellowship.

\newpage

\figcaption{
The net spectrum of RX~J185635-3754 observed with
the {\it EUVE} SW detector.
For comparison we show the blackbody (BB) fits to
{\it ROSAT} data alone (Table~\ref{tab2}; dashed curve) and to the
combined optical and X-ray data (Table~\ref{tbl-mwf}; solid curve).
\label{fig_euve}}  

\figcaption[fig_asca.ps]{
The background-subtracted count spectrum of
RX~J185635-3754 observed with {\it ASCA}.
The lower panel shows the combined SIS
spectrum; the combined GIS spectrum is in the upper panel.
\label{fig_asca}}   

\figcaption{Emergent (non-redshifted) spectral flux distributions for
selected compositions for $g_{14}=2.43$.
Curves are labelled by their
$\log T_{eff}$ and composition (BB -- dotted lines, Fe -- solid lines,
He -- dashed lines, and H -- dot-dashed lines).  The temperatures
chosen for the spectra in the left panel correspond to those of Figure 2
in \cite{RR96}, and those in the right panel correspond to those in Figure
5 of \cite{ZPS96}.  
\label{comp}}   

\figcaption{Emergent (non-redshifted) spectral flux distributions from
Fe atmospheres with $\log T_{eff} = 5.6$ and the indicated surface
gravities $g_{14}$.  The BB flux is also shown for
comparison. \label{fig_grav}}   

\figcaption {The best-fit Fe model (upper panel) and residuals (lower
panel) to the {\it ROSAT} PSPC data.  The parameters of the fit are
listed in Table \ref{tab2}.\label{fitFe} }   

\figcaption {The effects of varying the redshift on the best fits
to the {\it ROSAT} PSPC data.  Thick (thin) lines indicate
$T_\infty$ ($n_H$) for both Fe (dashed lines) and Si-ash (solid lines) models.
Errors (plotted) are about $\pm1.8$~eV for $T_\infty$ and $\pm$0.2 for
$n_{H,20}$, at a given $z$. For clarity, the errors on $n_{H,20}$ are
indicated only on the Si-ash points.
\label{fz1}}   

\figcaption {The correlations between $n_H$ and $T_\infty$ for the
best-fit Fe (dashed line) and Si-ash (solid line) models to the
{\it ROSAT} PSPC data at fixed values of the redshift.
 The redshift decreases from left to right.  For comparison, the
best-fit BB model is indicated by a diamond.
\label{fz2}}   

\figcaption {The spectral flux distribution in the 1 eV -- 1
keV range, for blackbody models.
The dashed curve is the best-fit model
to the {\it ROSAT} PSPC data, while the solid curve is the best-fit
model to the combined optical, {\it EUVE} and {\it ROSAT} data.
Both models are generated for $z$=0.305. The
{\it ROSAT} PSPC data are
not displayed because the energy redistribution of the response matrix
does not permit a model-independent flux to be determined.  The {\it EUVE}
data are denoted by the error bars in the 0.12--0.18 keV range. 
The mean STIS flux is shown; the spectrum is shown on a larger scale in
Figure~\ref{uvcons}. Error bars on all points 
denote the photometric uncertainties; horizontal bars denote the
band widths for the broadband photometry.
\label{figbb}}

\figcaption {Same as Figure \ref{figbb} but for a pure Fe atmosphere.
The dotted curve is the best multiwavelength fit at the optimal $z$.
\label{figFe}}

\figcaption
{Same as Figure \ref{figFe} but for a Si-ash atmosphere.
The dotted curve (barely distinguishable from the solid curve)
is the best multiwavelength fit at the optimal $z$.
\label{figSi}}

\figcaption {Same as Figure \ref{figbb} but for a pure H atmosphere.
Note the difference in scale.
In this case (and that for He, not shown), no acceptable joint fits of
optical, {\it EUVE} and {\it ROSAT} data are possible.
The absorption feature in the models near 0.08~keV is H~I Lyman~$\alpha$
for $z$=0.305.
\label{figH}}

\figcaption {The acceptable regions in the $R_\infty/D$ -- $kT_\infty$
plane for
the blackbody fits. Three~$\sigma$ confidence contours are drawn.
The thick contour denotes the best fit to the multiwavelength data, with the
four data sets given equal weight. The
best fits to the individual data sets are also shown. Contours of constant
n$_H$ (=1, 2, and 3$\times$10$^{20}$~cm$^{-2}$ from bottom to top)
are overplotted. Note that the regions allowed by the X-ray data follow
the n$_H$ contours, while the optical and UV data, taken longward of the peak
of the emission, allow regions of approximately constant $R_\infty/D$. The
thick dots mark the limits of the region explored. Note that at 
3$\sigma$ confidence the regions allowed by the PSPC and the optical/UV data
do not intersect.
\label{fig-bbgrid}}

\figcaption {Same as Figure \ref{fig-bbgrid} but for a pure Fe atmosphere.
At 3$\sigma$ confidence the regions allowed by the PSPC and the optical/UV data
do not intersect, but the discrepancy is less than for the blackbody model.
\label{fig-Fegrid}}

\figcaption
{Same as Figure \ref{fig-bbgrid} but for a Si-ash atmosphere.
At 3$\sigma$ confidence the regions allowed by the PSPC and the optical/UV data
do intersect.
\label{fig-Sigrid}}

\figcaption {Same as Figure \ref{fig-bbgrid} but for a pure H atmosphere.
The X-ray and optical/UV regions cannot be reconciled. For clarity, we plot
4$\sigma$ confidence contours. The best formal fit
region lies near the optical 4~$\sigma$ confidence contour, at
$kT_\infty$=26~eV.
\label{fig-Hgrid}}

\figcaption {The effects of redshift on the derived value of
$R_\infty/D$, for the Fe (right panel) and Si-ash (left panel) models.
The solid curves are obtained by fitting the {\it ROSAT} PSPC data with a
uniform-temperature model, and the filled bands are determined from
the optical observational constraint (\ref{fopt}).  Their intersections
denote the combined acceptable values of $z$ and $R_\infty/D$.
In each case, the dashed lines
bracketing the solid lines, and the width of the shaded bands,
indicate the uncertainties in $R_\infty/D$.\label{fz3}}

\figcaption {Mass-radius diagrams for the uniform-temperature 
heavy element atmosphere models.
Upper and lower panels are for Fe and Si-ash compositions,
respectively.  Solid and dashed curves are for equations of state
labelled following Lattimer \& Prakash (2001).  The dashed line
labelled ``causality'' is the compactness limit set by requiring
equations of state to be causal.  Dotted lines are contours of fixed
$R_\infty$.  The crosses denote masses and radii of models which best
fit the optical and X-ray data, for the indicated distances, and the
hatched regions surrounding them include the nominal errors indicated
in the constraint relations \ref{nhz}--\ref{fopt}.
\label{mr1}}

\figcaption {The spectral flux distribution for a two-component BB
model with $T_{H\infty}=55.3$~eV and $T_{C\infty}=20$~eV.
The dashed
line indicates the best-fit uniform-temperature BB model ($\alpha=1$),
and the solid curve is the best 2-component component fit, with $\alpha=0.22$.
The cool component makes no significant contribution to the X-ray flux.
Other notation is the same as in Figures \ref{figbb}-\ref{figH}.
\label{twobb}}

\figcaption {The effective increase in $R_\infty$ that can be obtained
by progressively lowering $T_{C\infty}$ in the two-component BB model
fit to the combined optical and X-ray data.  The thick solid curve is for
the assumed 61~pc distance, and dashed curves indicate \ud{9}{8}~pc
deviations from this.  The thin solid curve shows $\alpha$ as a function
of $T_{C\infty}$.
$T_{H\infty}$ is fixed to 55.3~eV, the value
for the best-fit uniform-temperature BB fit.  
The vertical dotted line indicates the lower limit to $T_{C\infty}$,
15~eV, from the slope of the optical and UV fluxes. This formulation 
assumes that the cool component does not contribute any significant flux 
at X-ray wavelengths, and so breaks
down as $T_{C\infty}$ approaches $T_{H\infty}$.
\label{2bbc}}

\figcaption{The far UV spectrum from STIS with $\pm1\sigma$ envelope
contours (dotted lines). The data have been smoothed with a Fourier filter
and a 7-pixel (4.1\AA) running mean.
Uncertainties are particularly
large in the vicinity of the subtracted geocoronal H I Lyman-$\alpha$ and O I
emission at 1216 and 1300 \AA, respectively.
There are no statistically significant emission or absorption features in
this spectrum.
The lower solid curve is the baseline BB fit
($T_\infty=55.3$~eV, $R_\infty$/D=0.070~km~pc$^{-1}$).
The upper three
curves are BB fits to the residual flux above the baseline for
$T_\infty$= 5, 10, and 25~eV, respectively. The inset shows the
optical-UV spectral energy distribution on a  log-log plot with flux
units identical to that of the main plot. The crosses show the mean 
STIS flux and the mean F170W, F300W, F450W, and F606W fluxes, together with
the relevant wavelength ranges.  The smooth curves are the extrapolations
of the $T_\infty$= 5, 10, and 25~eV BB curves to longer wavelengths.
\label{uvcons}}

\figcaption{Contours of $\alpha$ as a function of the gravitational redshift
$z$ and the temperature of the cool component $T_{C,\infty}$. The vertical
dotted line represents the value of $z$ below which there is an optical
excess. The horizontal dashed lined, at $T_{C,\infty}$=22~eV, is the coolest
permissable temperature based on the optical spectrum. The diagonal
dashed line is at half of $T_{H,\infty}$, and represents approximately
the point above which the cool component begins to noticeably affect the
shape of the high energy spectrum. Allowable regions lie within
the wedge between the dashed lines. Kinks in the curves are a consequence of the
discrete grid used.
\label{fig_alphaz}}

\figcaption{$R_\infty$/D as a function of $z$ for the two-component
heavy element atmospheres. 
The vertical dotted lines represent the lower limits for $z$; below this the
model predicts too much optical emission.
Addition of a cool component can only increase $R_\infty$/D
by up to about 60\% over the single temperature component value, and then
only for very high values of $z$.
\label{fig_maxrad}}

\figcaption{Mass as a function of $z$ for the two-component heavy element
atmospheres. Masses are plotted only for points for redshifts for which the
two-component model is valid.
\label{fig_az3}}

\figcaption{Similar to figure \ref{mr1}, but for two-temperature fits to
X-ray and optical data.  The shaded region is the 90\% confidence
region allowed by the
constraints (\ref{fopt2})--(\ref{fthc}),
for an assumed distance of 61~pc.
\label{mr3}}

\figcaption{Cooling curve observations for neutron stars for which
thermal emission is believed to be seen.  Except for RX~J185635-3754, data
are from Pavlov (2001; private communication)
and ages are the standard pulsar spin-down times,
or the remnant age if known. In many cases, luminosities derived from
different assumed compositions are shown with errors (mostly resulting
from distance uncertainties). MH indicates magnetic hydrohen atmospheres; BB
indicates blackbody fits.
The luminosity for RX~J185635-3754 is 
that of the heavy element atmospheres; the two-component blackbody model
has a similar luminosity.
  \label{lumrxj}} \clearpage


\begin{deluxetable}{rrrrrrrrrr}
\tablecolumns{10}
\tablecaption{{\it ROSAT} Observation Log}
\tablehead{
\colhead{ROR} & \colhead{Instrument} & \multicolumn{4}{c}{Start}
   & \multicolumn{3}{c}{Stop} &  \colhead{Length}\\
    && \multicolumn{7}{c}{(UT)} & \colhead{(ksec)} } 
\startdata
200497 & PSPC & 1992 & 11 & Oct & 10:06 & 16 & Oct & 5:33 &  6.3\\
400612 & HRI  & 1994 &  7 & Oct & 13:06 &  8 & Oct &22:22 & 17.8\\
400864 & HRI  & 1997 &  9 & Oct & 14:34 & 15 & Oct & 9:39 & 29.1\\
\enddata
\label{tbl-rosobs}
\end{deluxetable}

\begin{deluxetable}{rrr}
\tablecolumns{3}
\tablecaption{{\it EUVE} Observing Log}
\tablehead{
\colhead{Start} & \colhead{Stop} & \colhead{Length}\\
\multicolumn{2}{c}{(1997 UT)} & \colhead{(ksec)} }
\startdata
18 June 04:26 &  24 Jun 23:29 &  195\\
26 June 09:59 &  28 Jun 22:07 &   60\\
30 June 22:24 &   3 Jul 12:42 &   77\\
 9 July 00:33 &  16 Jul 00:23 &  180\\
16 July 21:44 &  19 Jul 08:50 &   90\\
\enddata
\label{tbl-euve}
\end{deluxetable}

\clearpage

\begin{deluxetable}{lrrlrrr}
\tablecolumns{7}
\tablecaption{{\it HST} Observations}
\tablehead{
\colhead{Root} &
\colhead{Filter} & \colhead{exposure} & \multicolumn{2}{c}{Start} &
     \colhead{f$_\lambda$} & \colhead{$\sigma$}\\
     & &  \colhead{(sec)} & \multicolumn{2}{c}{(UT)} &
      \colhead{(erg cm$^2$ s$^{-1}$ \AA$^{-1}$)}
       }
\startdata
U3IM010x, x=1-4 & F606W &  4400 & 1996 Sep 30 & 2:25 & 1.29$\times$10$^{-19}$ & 16.7\\
U51G010x, x=1-8 & F606W &  7200 & 1999 Mar 30 & 0:35 & 1.51$\times$10$^{-19}$ & 23.2\\
U51G020x, x=3-6 & F606W &  5190 & 1999 Sep 16 & 7:51 & 1.42$\times$10$^{-19}$ & 19.2\\
U625010x, x=1-6 & F606W &  7400 & 2001 Mar 24 &19:32 & 1.37$\times$10$^{-19}$ & 22.0\\
U51G040x, x=1-8 & F450W &  7200 & 1999 May 26 &17:20 & 4.26$\times$10$^{-19}$ & 15.3\\
U3IM010x, x=5-6 & F300W &  2400 & 1996 Oct 12 & 5:29 & 1.90$\times$10$^{-18}$ & 7.3\\
U51G020x, x=1-2 & F300W &  2600 & 1999 Sep 16 & 6:15 & 2.48$\times$10$^{-18}$ & 6.9 \\
U51G030x, x=1-4 & F170W & 10800 & 1999 May 24 &16:43 & 1.45$\times$10$^{-17}$ & 5.8\\
\enddata
\label{tbl-hst}
\end{deluxetable}

\clearpage

\begin{deluxetable}{lccccccccc} 
\tablecolumns{10}
\tablecaption{Si-ash Composition\tablenotemark{a}\label{tab1}}
\tablehead{
\colhead{Element} & \colhead{Si} & \colhead{Ar} & \colhead{Ti} &
   \colhead{Mn} & \colhead{Ni} & \colhead{S} & \colhead{Ca} & 
   \colhead{Cr} & \colhead{Fe} }
\startdata
Abundance (\%) & 11 & 3 & 1 & 1 & 2 & 10 & 3 & 1 & 68 
\enddata
\tablenotetext{a}{from Arnett (1996)}
\end{deluxetable}

\begin{deluxetable}{cccccccc}
\tablecolumns{8}
\tablecaption{{\it ROSAT} PSPC Spectral Fits\label{tab2}}
\tablehead{
   \colhead{Model} & \colhead{$z$} &
   \colhead{$n_H$} & \colhead{$T_{\infty}$} &
   \colhead{$\chi^2_\nu$\tablenotemark{a}} &
   \colhead{$R_\infty/D$} & \colhead{$D$\tablenotemark{b}} &
   \colhead{$R_\infty$\tablenotemark{c}} \\
    & & \colhead{($10^{20}$ cm$^{-2}$)}
    & \colhead{(eV)} & & \colhead{(km pc$^{-1}$)}
    & \colhead{(pc)} & \colhead{(km)}
}
\startdata
\cutinhead{Fits with $z$ fixed at the nominal value of 0.305}
BB     & \nodata & 1.73 $\pm$ 0.13 & 55.3 $\pm$ 5.5 & 1.43 & 0.070 $\pm$ 0.015 &
         186 $\pm$ 40 & 4.3 $\pm$1.1 \\
H      & \nodata & 2.36 $\pm$ 0.10 & 13.2 $\pm$ 1.6 & 1.44 & 2.19 $\pm$ 0.57  & 
         6.0 $\pm$ 1.6 & 134 $\pm$35 \\
He     & \nodata & 2.40 $\pm$ 0.10 & 13.3 $\pm$ 1.6 & 1.44 & 2.38 $\pm$ 0.60  & 
         5.5 $\pm$ 1.4 & 145 $\pm$ 37 \\
Fe     & \nodata & 2.07 $\pm$ 0.14 & 33.7 $\pm$ 1.5 & 1.86 & 0.290 $\pm$ 0.047 
       & 45 $\pm$ 7 & 18 $\pm$ 3 \\
Si-ash & \nodata & 1.88 $\pm$ 0.11 & 42.6 $\pm$ 1.6 & 1.72 & 0.147 $\pm$ 0.020 &
         89 $\pm$ 12 & 9.0 $\pm$ 1.2 \\
\cutinhead{Best fits with $z$ allowed to vary}
H      &0.43 & 2.44 $\pm$ 0.10 & 12.7 $\pm$ 1.5 & 1.43 & 2.52 $\pm$ 0.62  & 
            5.2 $\pm$ 1.3 & 154 $\pm$ 38 \\
Fe     &0.64 & 0.96 $\pm$ 0.14 & 80.0 $\pm$ 3.5 & 1.49 & 0.028 $\pm$ 0.002 & 
         466 $\pm$ 33 & 1.7 $\pm$ 0.1\\
Si-ash &0.54 & 1.17 $\pm$ 0.16 & 69.1 $\pm$ 4.0 & 1.45 & 0.036 $\pm$ 0.005 &
         363 $\pm$ 50 & 2.2 $\pm$ 0.3 \\

\enddata
\tablenotetext{a}{Reduced $\chi^2$ for 86 degrees of freedom.}
\tablenotetext{b}{Distance assuming $R_\infty$=13.05 km.} 
\tablenotetext{c}{$R_\infty$ assuming distance = 61\ud{9}{8}~pc.} 
\end{deluxetable}

\begin{deluxetable}{ccccccc}
\tablecolumns{7}
\tablecaption{Parameters from Multiwavelength
              Fits\tablenotemark{a}\label{tbl-mwf}}
\tablehead{
   \colhead{Model}        & \colhead{$n_H$} &
   \colhead{$T_{\infty}$} & \colhead{$R_\infty/D$} &
   \colhead{$T_\infty(R_\infty/D)^2$} & \colhead{Luminosity} &
   \colhead{P$_{OX}$\tablenotemark{b}}    \\
    & \colhead{($10^{20}$ cm$^{-2}$)} &
   \colhead{(eV)} & \colhead{(km pc$^{-1}$)} &
   \colhead{(eV (km pc$^{-1}$)$^2$)} &
   \colhead{(10$^{31}$ erg s$^{-1}$)\tablenotemark{c}}
    }

\startdata
BB     & 2.2\ud{0.3}{0.4} & 48$\pm$2 & 0.11$\pm$0.01 & 
               0.60\ud{0.05}{0.4} & 1.55\ud{0.23}{0.17} & 3$\times$10$^{-4}$\\
H      & 1.0$\pm$0.1 & 26$\pm$1 & 0.27$\pm$0.01 &
                       1.94$\pm$0.01 & 0.6$\pm$0.01 & $<$10$^{-14}$ \\
Fe     & 1.8$\pm$0.2 & 44$\pm$1 & 0.13$\pm$0.01 & 
                   0.75$\pm$0.05 & 1.41\ud{0.08}{0.06} & 7$\times$10$^{-7}$\\
Si-ash & 1.9\ud{0.3}{0.2} & 45\ud{2}{1} & 0.13$\pm$0.01 &
                       0.74\ud{0.04}{0.05} & 1.63\ud{0.14}{0.21} & 0.53 \\

\enddata
\tablenotetext{a}{3$\sigma$ ranges, assuming $z$=0.305. Weighting of the data
              is discussed in the text.} 
\tablenotetext{b}{The likelihood that the X-ray and optical parameters are
                  the same.}
\tablenotetext{c}{Uncertainty does not include uncertainty in distance.}
\end{deluxetable}

\clearpage

\begin{deluxetable}{lccl}
\tablecolumns{4}
\tablecaption{Parameter Constraints\label{tbl-params}}
\tablehead{
   \colhead{Parameter}        & \colhead{Analytic Fit} &
   \colhead{Joint Acceptable Region} &
   \colhead{units}
    }

\startdata
\cutinhead{Si-ash model}
 $n_H$ & 1.84 $\pm$ 0.07  & 1.92 $\pm$0.05 &  10$^{20}$ cm$^{-2}$ \\
 $z$   & 0.31 $\pm$ 0.12  & 0.34 \ud{0.01}{0.03} & \\
 T     & 45.2 $\pm$  6.0  & 45 $\pm$1 & eV \\
 $R_\infty/D$   & 0.128 $\pm$ 0.021& 0.132 \ud{.011}{.008} & km pc$^{-1}$ \\

\cutinhead{Fe model}
 $n_H$ & 1.66 $\pm$ 0.11  & 1.81 $\pm$0.06 &  10$^{20}$ cm$^{-2}$ \\
 $z$   & 0.38 $\pm$ 0.12  & 0.37 \ud{0.04}{0.03} & \\
 T     & 44.3 $\pm$ 3.8   & 44.0 $\pm$1 & eV \\
 $R_\infty/D$   & 0.134 $\pm$ 0.021& 0.139 \ud{.010}{.012} & km pc$^{-1}$ \\

\enddata
\end{deluxetable}

\clearpage


\begin{figure}
\plotone{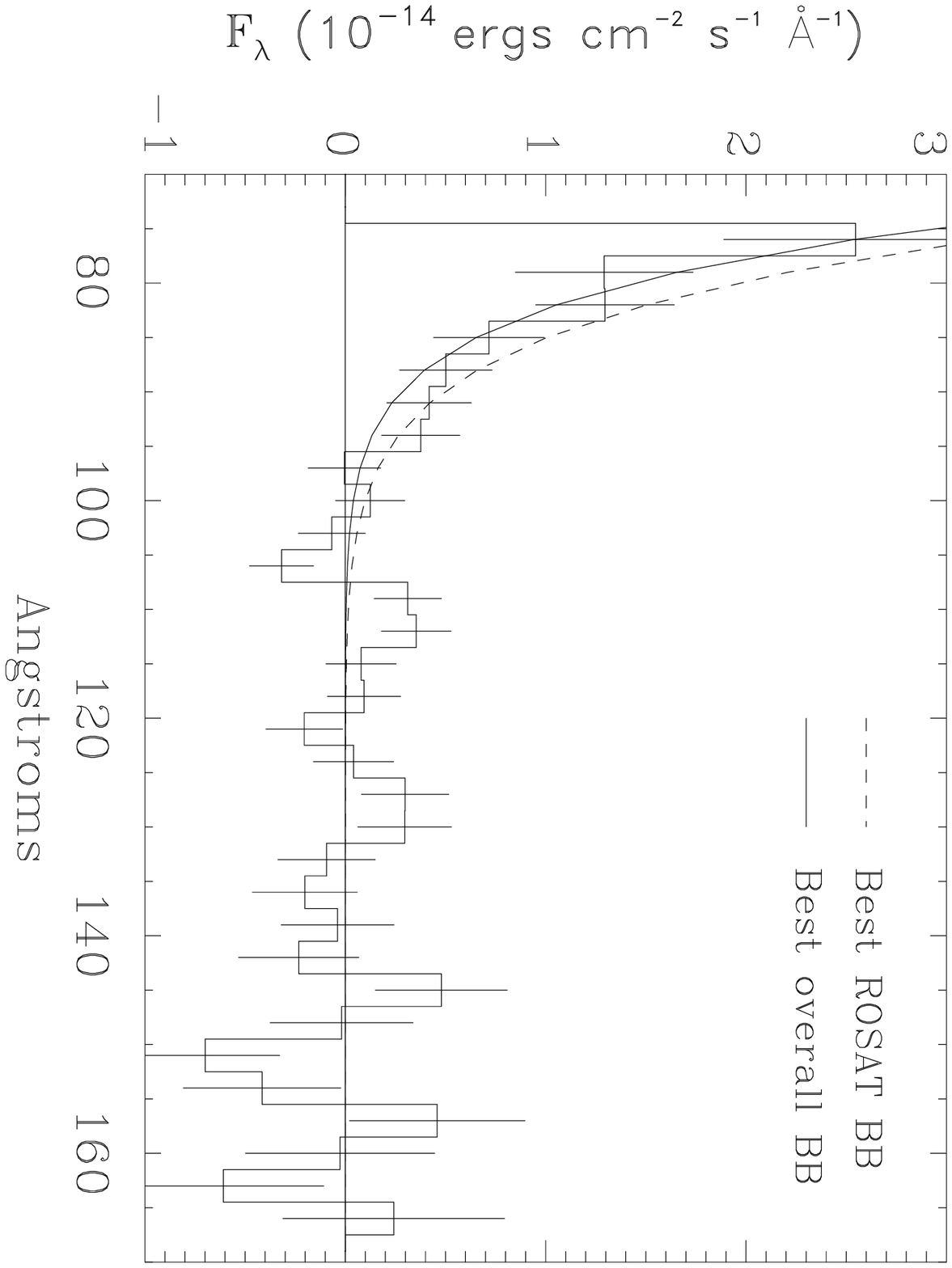}
\end{figure}

\begin{figure}
\plotone{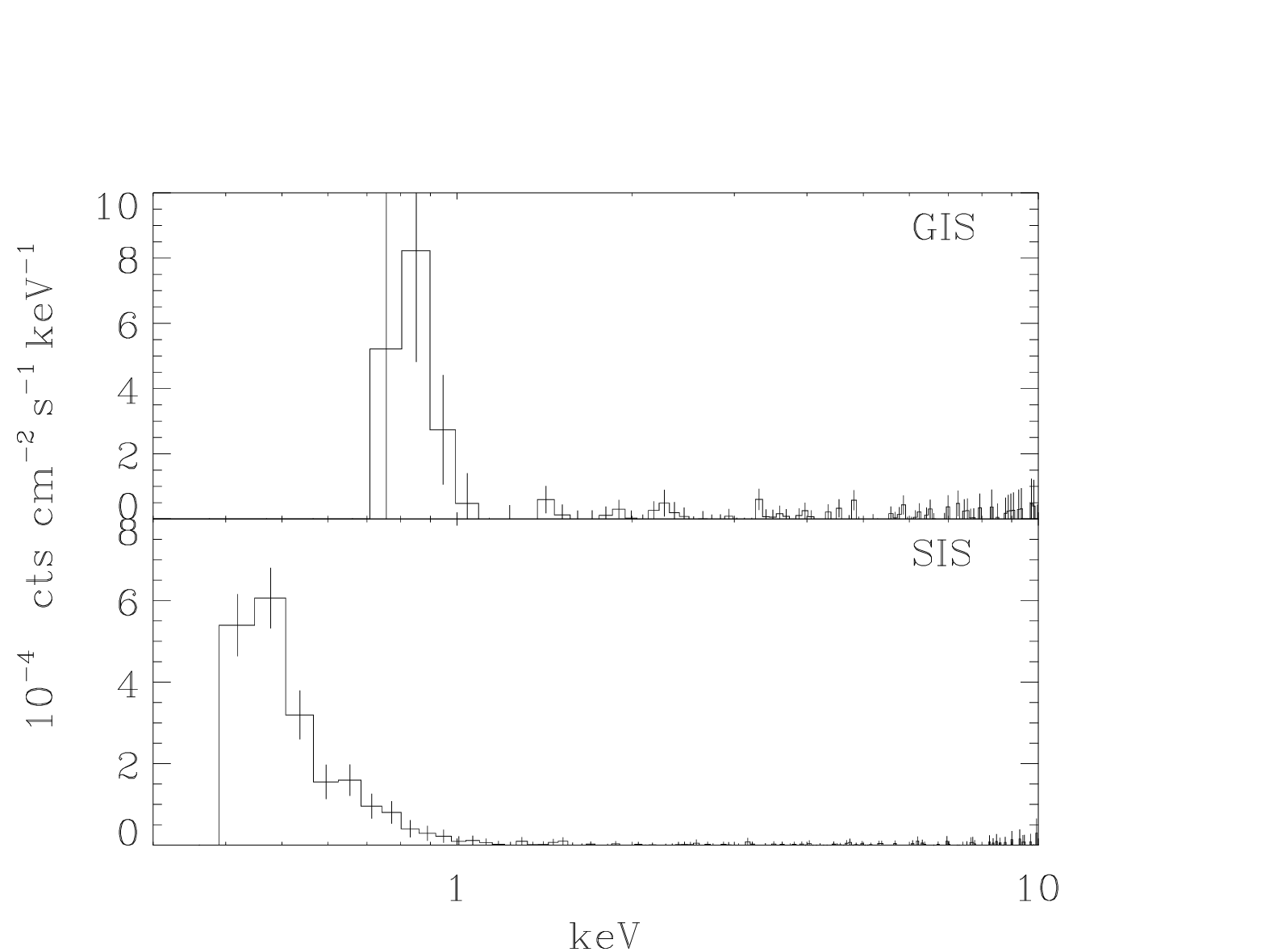}
\end{figure}

\begin{figure}
\epsscale{0.8}
\plotone{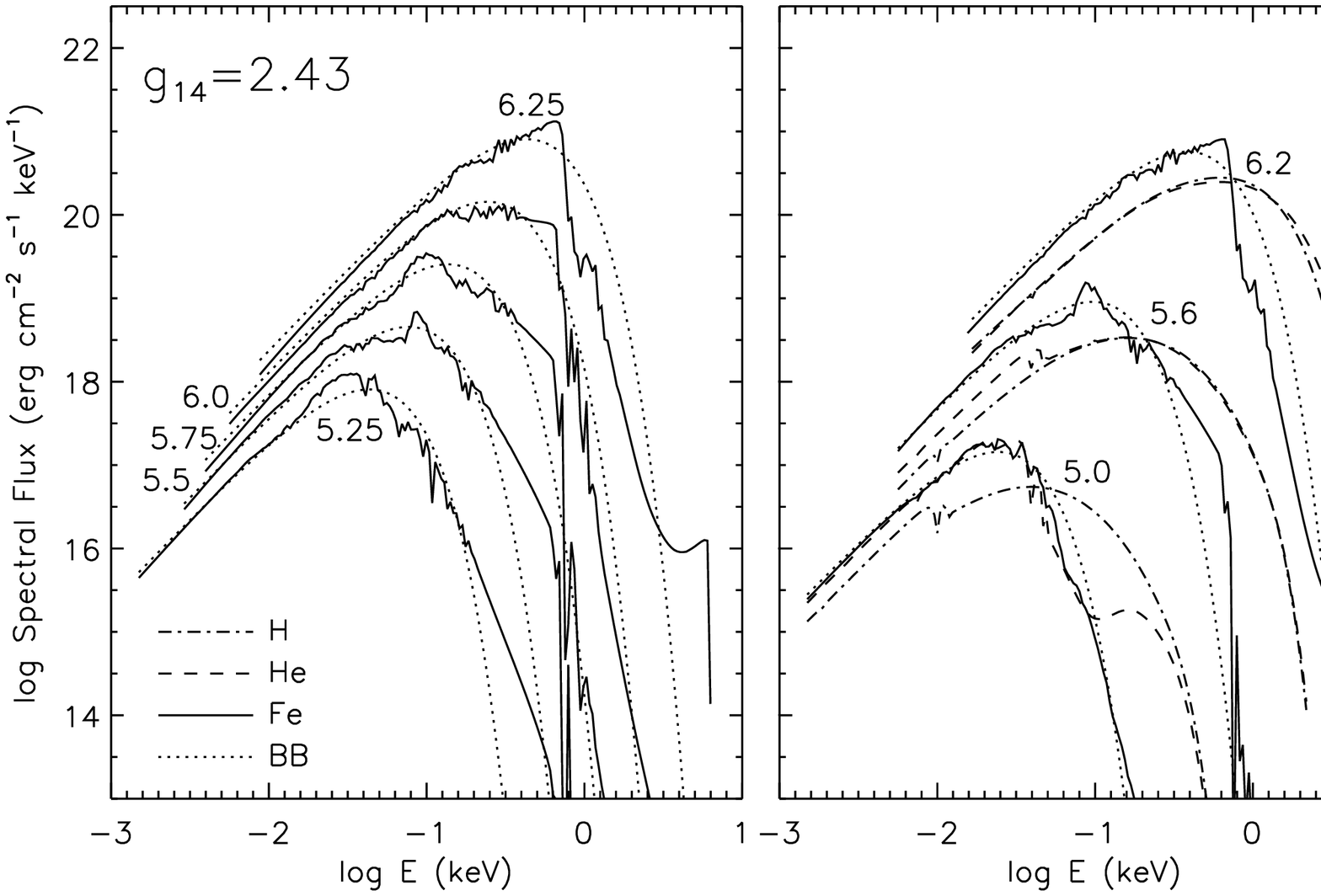}
\end{figure}

\begin{figure}
\epsscale{0.8}
\plotone{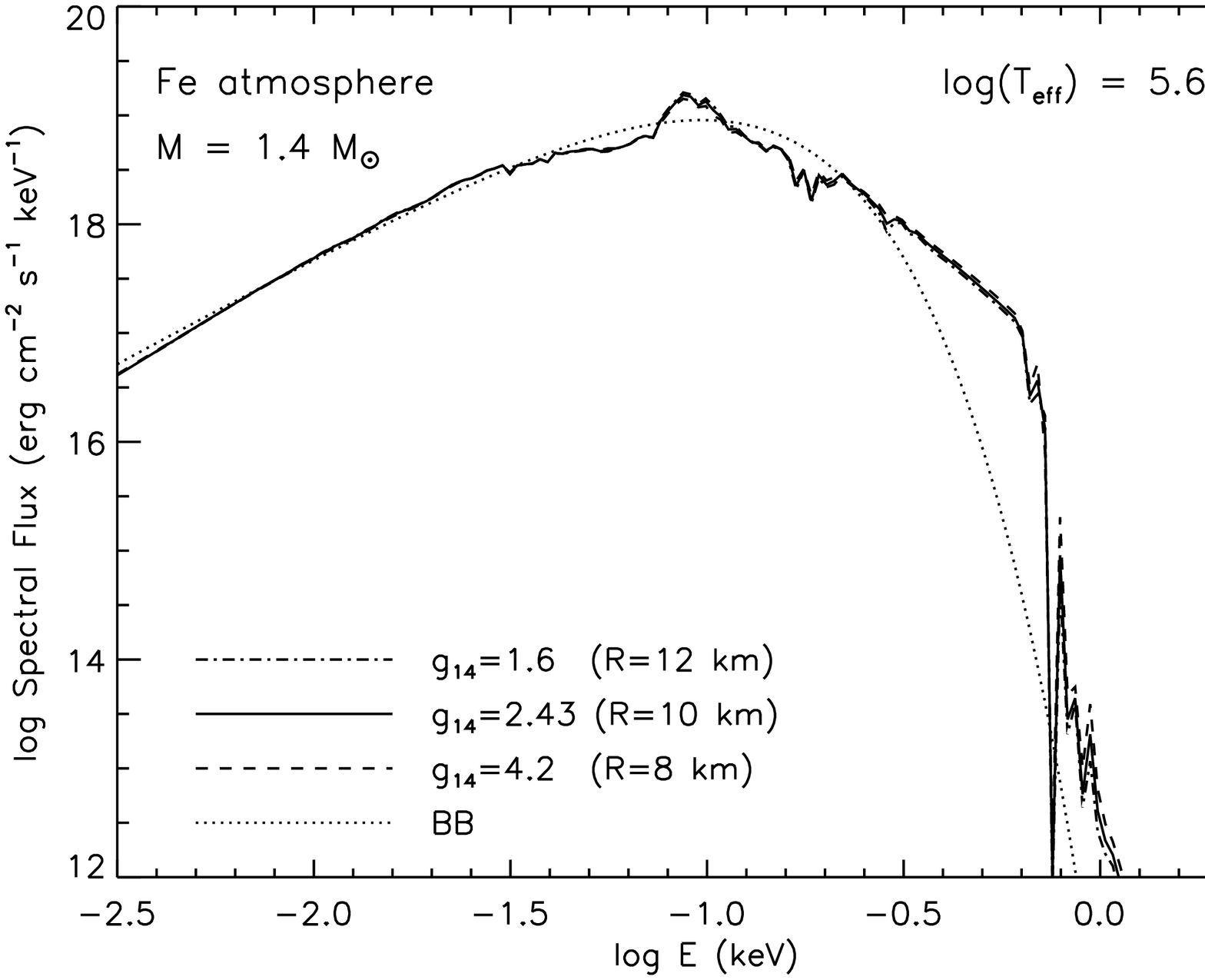}
\end{figure}

\begin{figure}
\epsscale{0.8}
\plotone{f5.ps}     
\end{figure}

\clearpage

\begin{figure}               
\epsscale{1.0}
\plotone{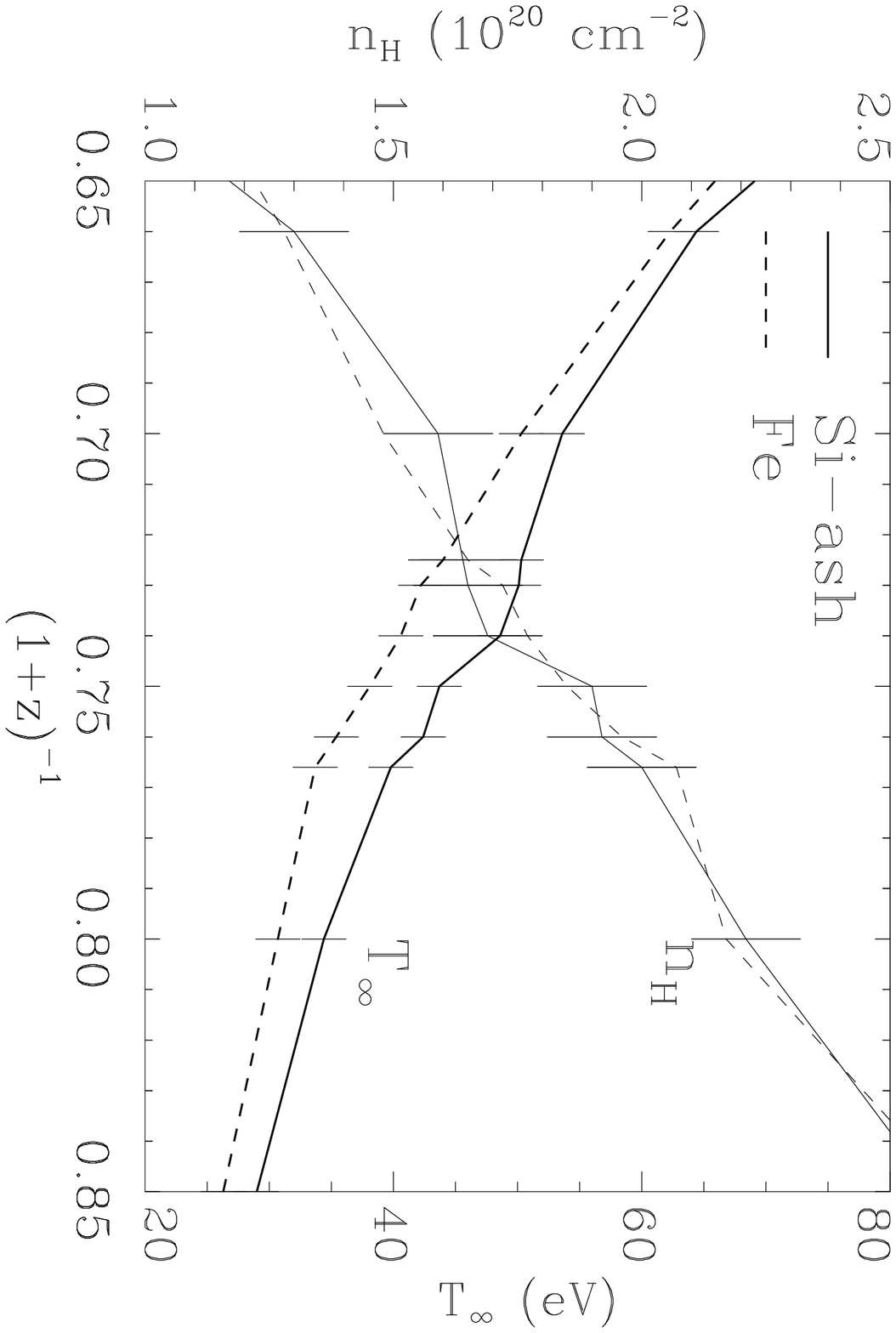}
\end{figure}

\begin{figure}               
\epsscale{1.0}
\plotone{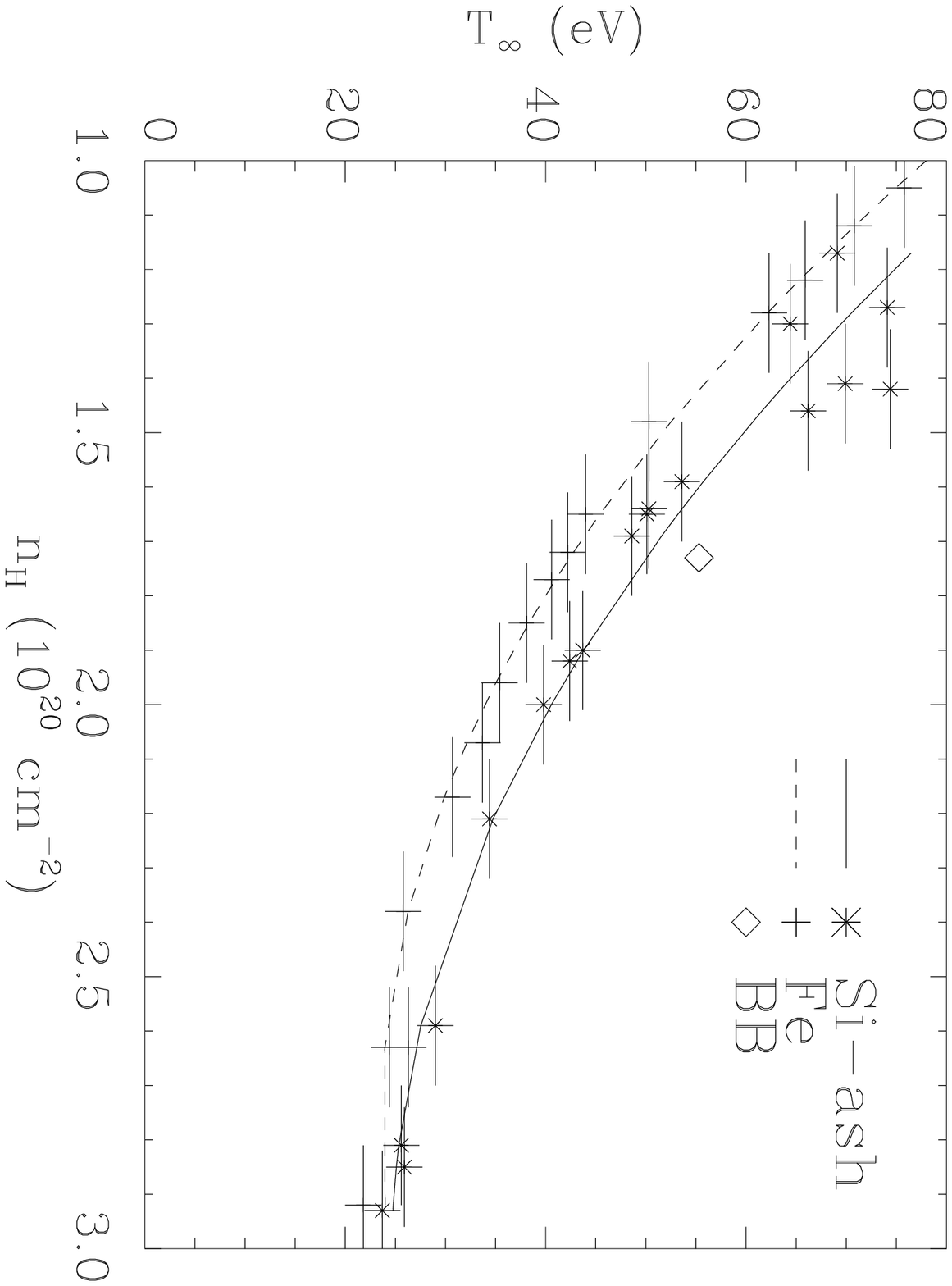}
\end{figure}

\begin{figure}     
\epsscale{0.8}
\plotone{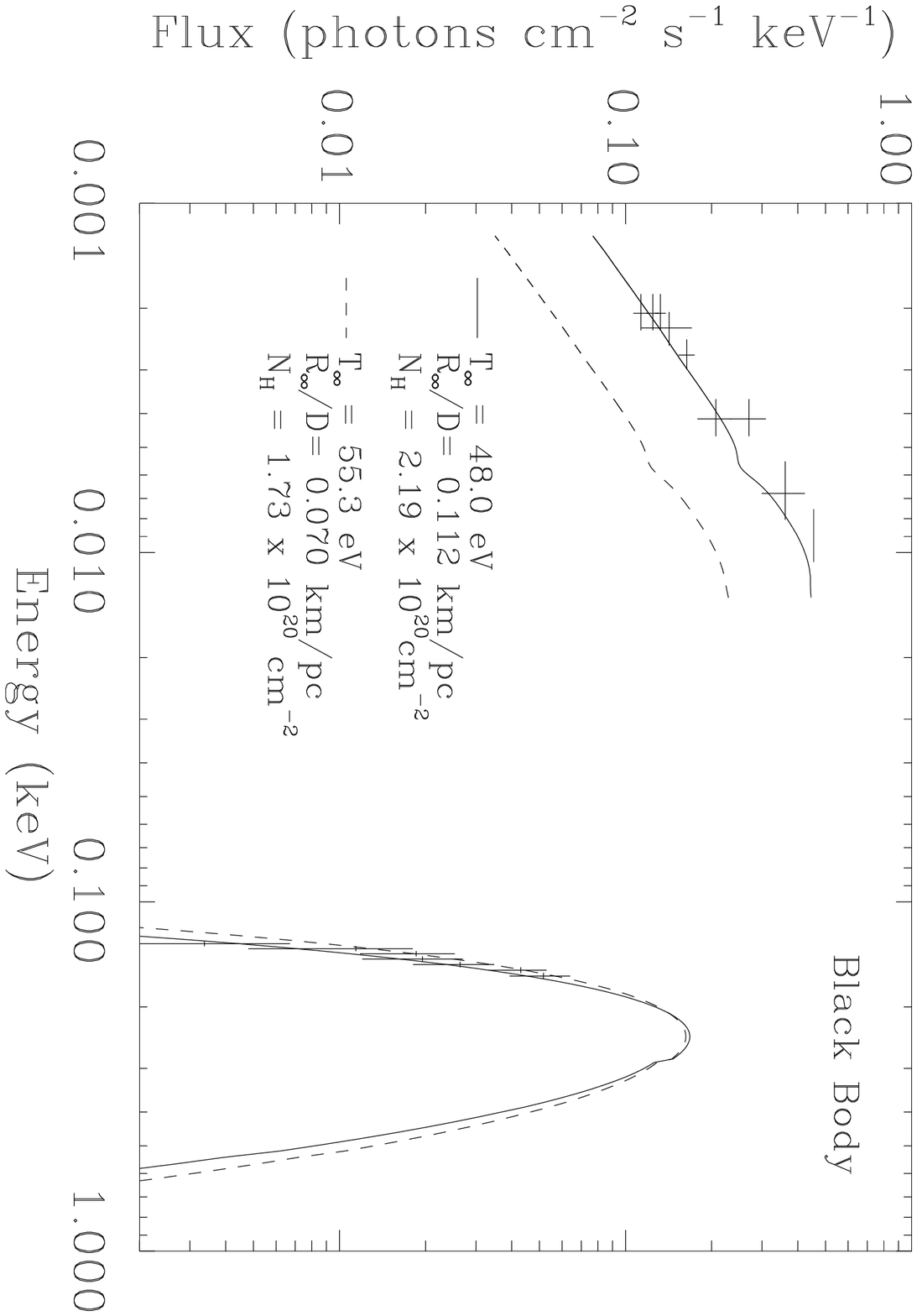}
\end{figure}

\begin{figure}
\epsscale{0.8}
\plotone{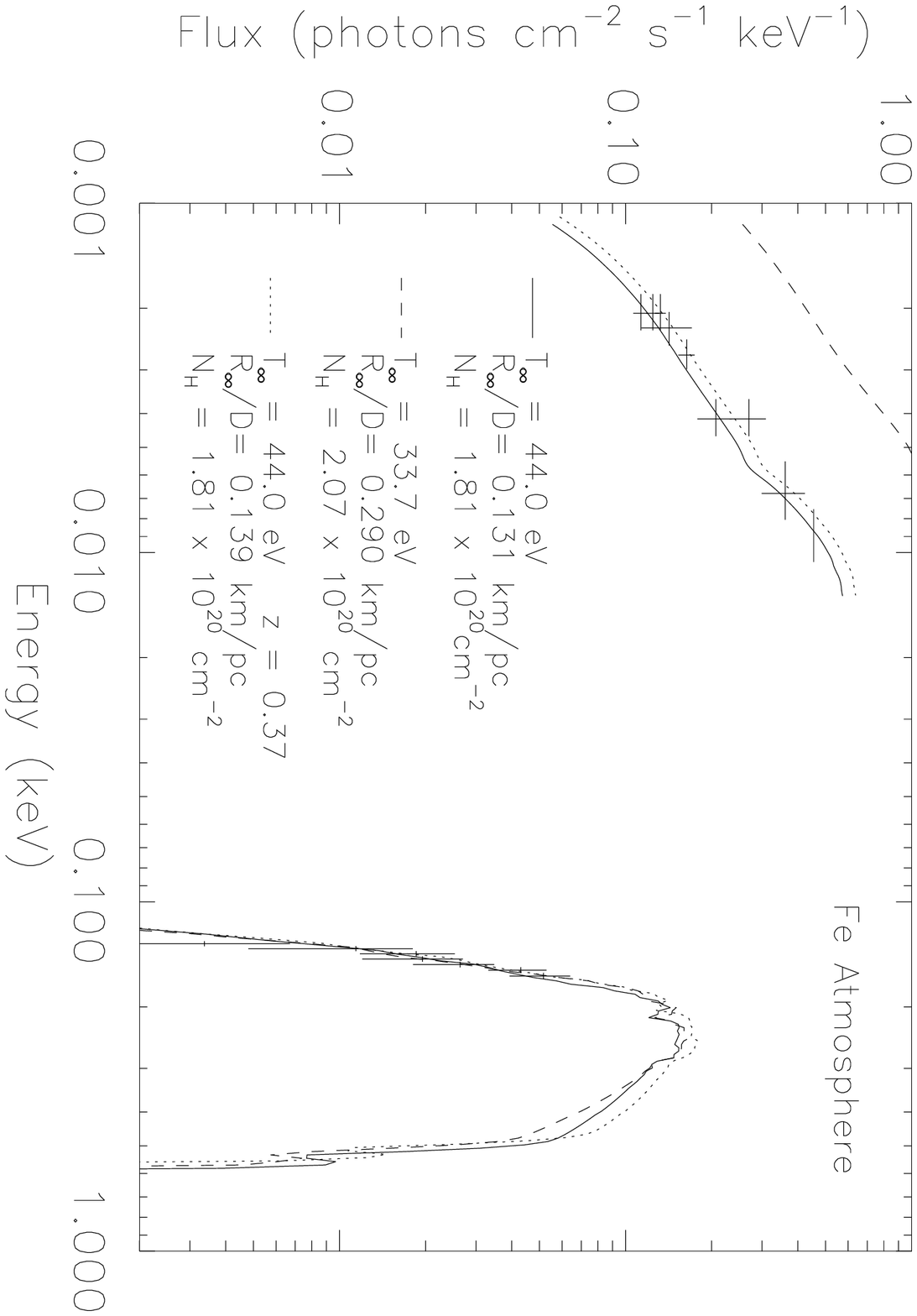}
\end{figure}

\begin{figure}
\epsscale{0.8}
\plotone{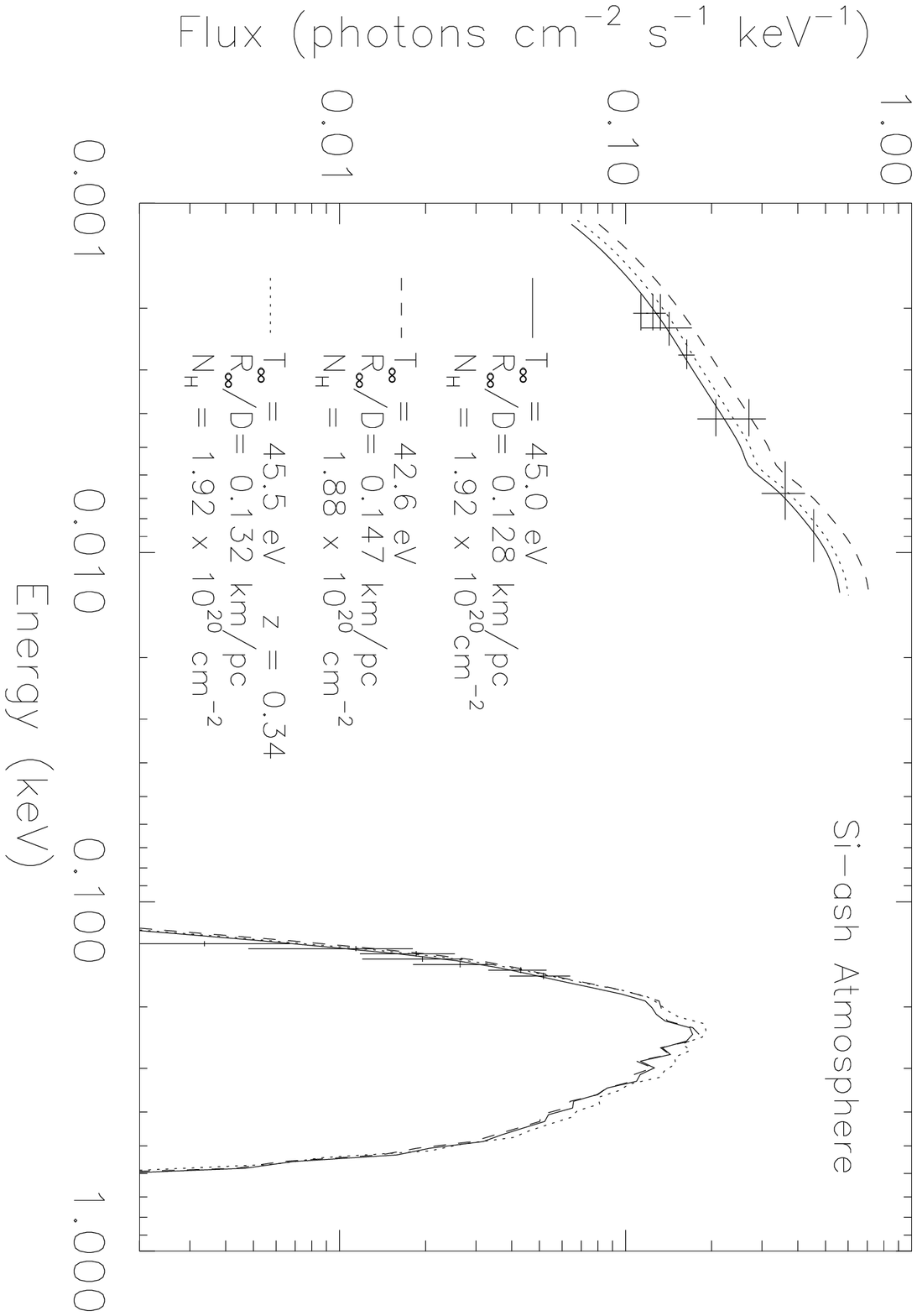}
\end{figure}

\clearpage

\begin{figure}               
\epsscale{0.8}
\plotone{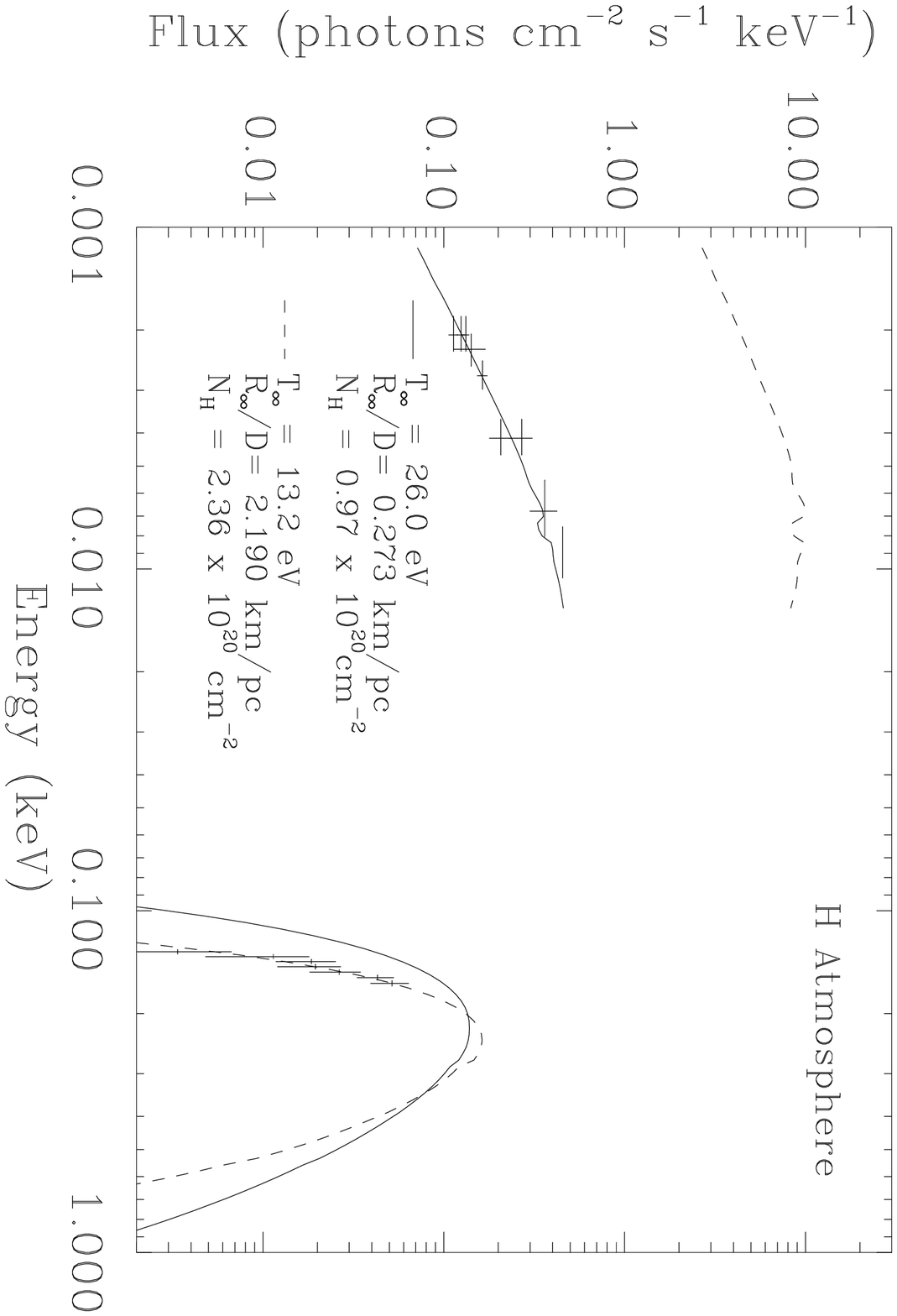}
\end{figure}

\begin{figure}               
\epsscale{0.8}
\plotone{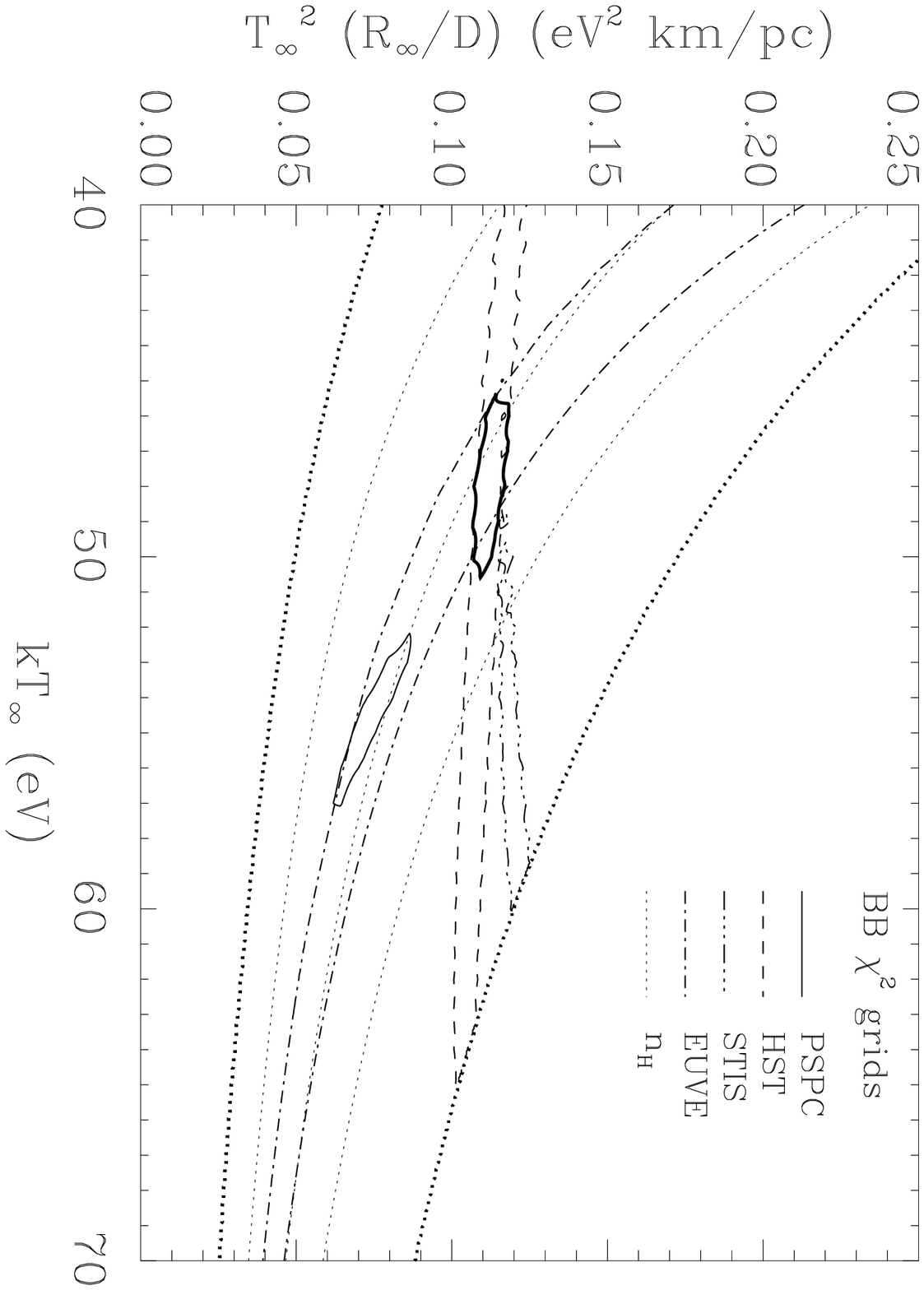}
\end{figure}

\begin{figure}               
\epsscale{0.8}
\plotone{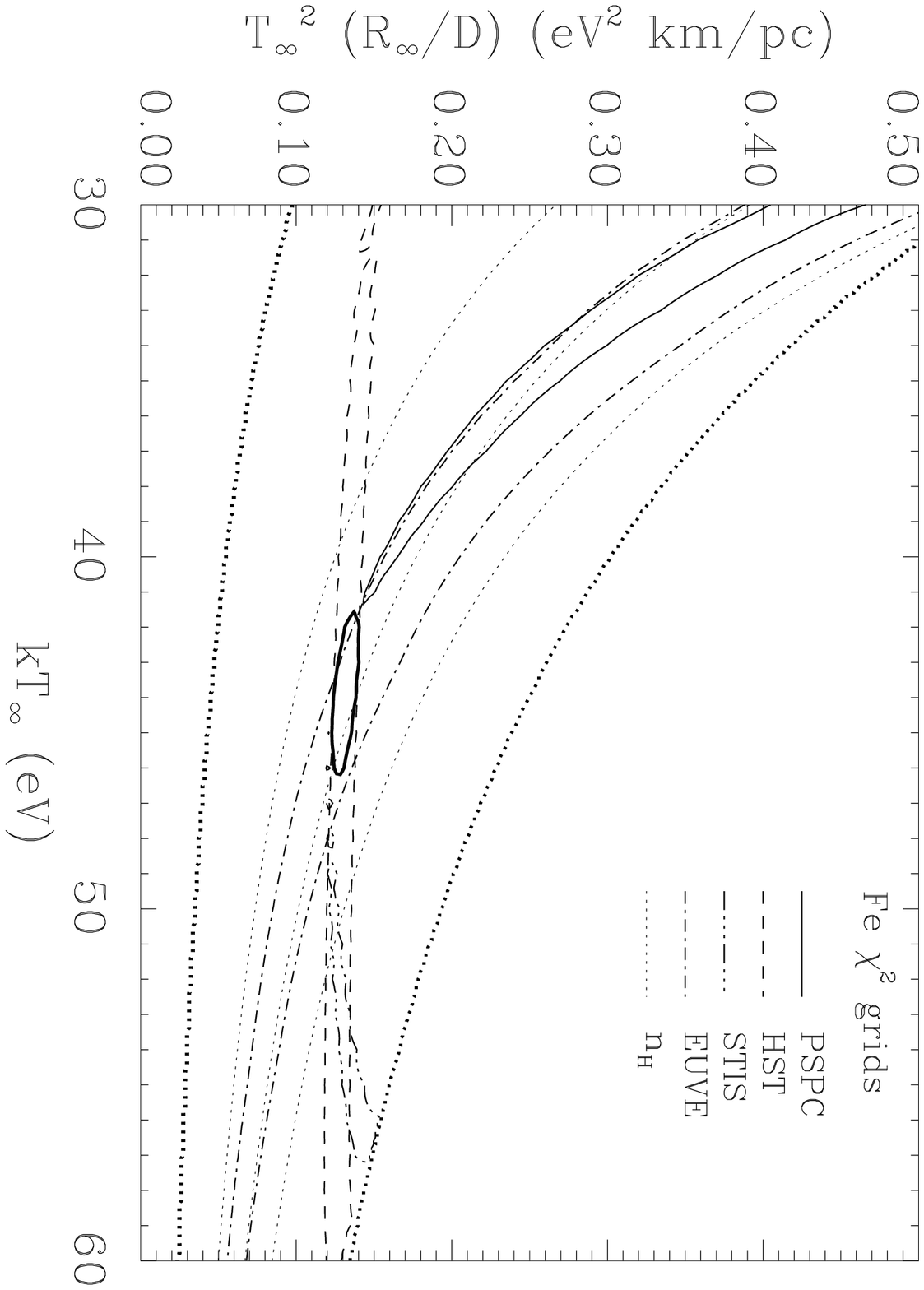}
\end{figure}

\begin{figure}               
\epsscale{0.8}
\plotone{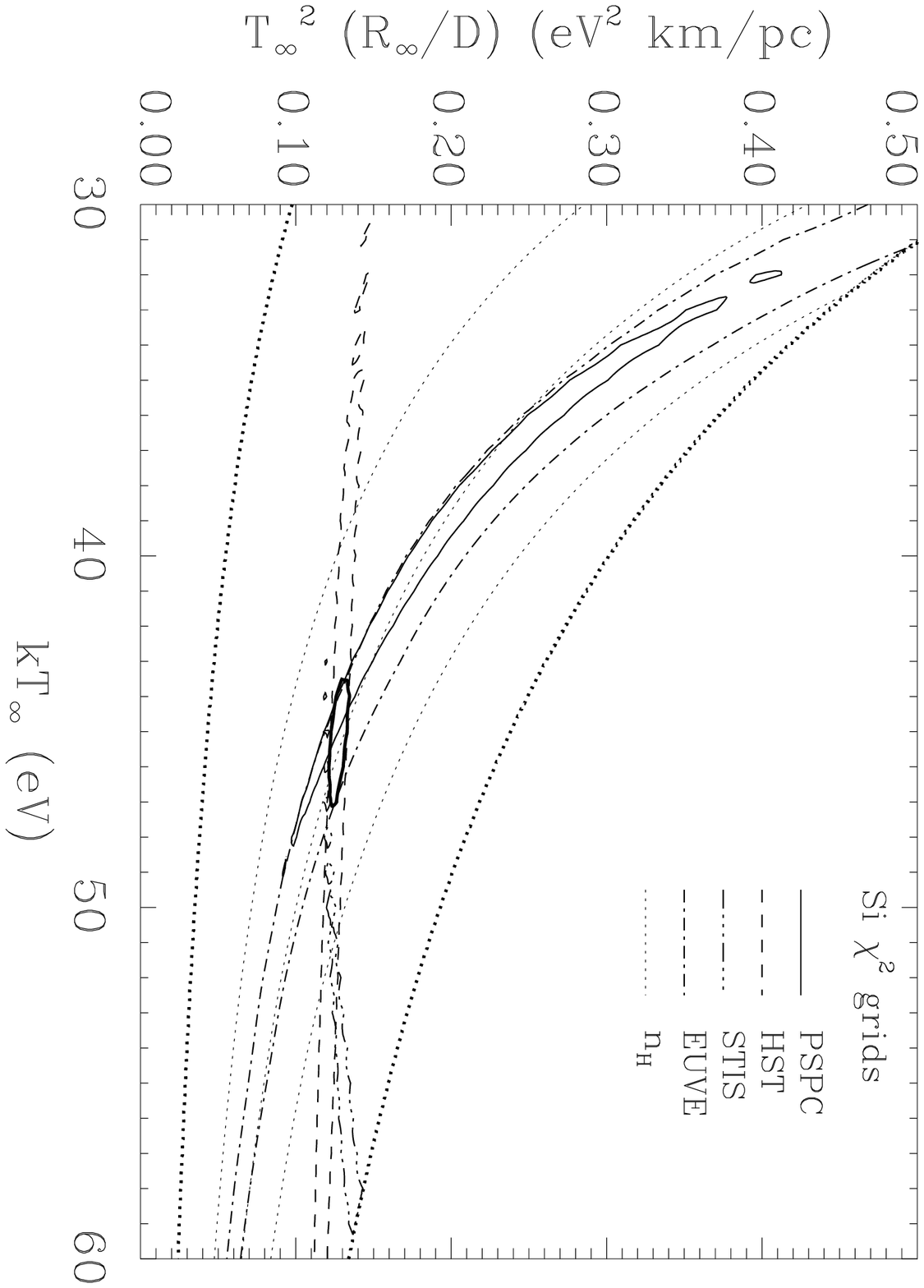}
\end{figure}

\begin{figure}               
\epsscale{0.8}
\plotone{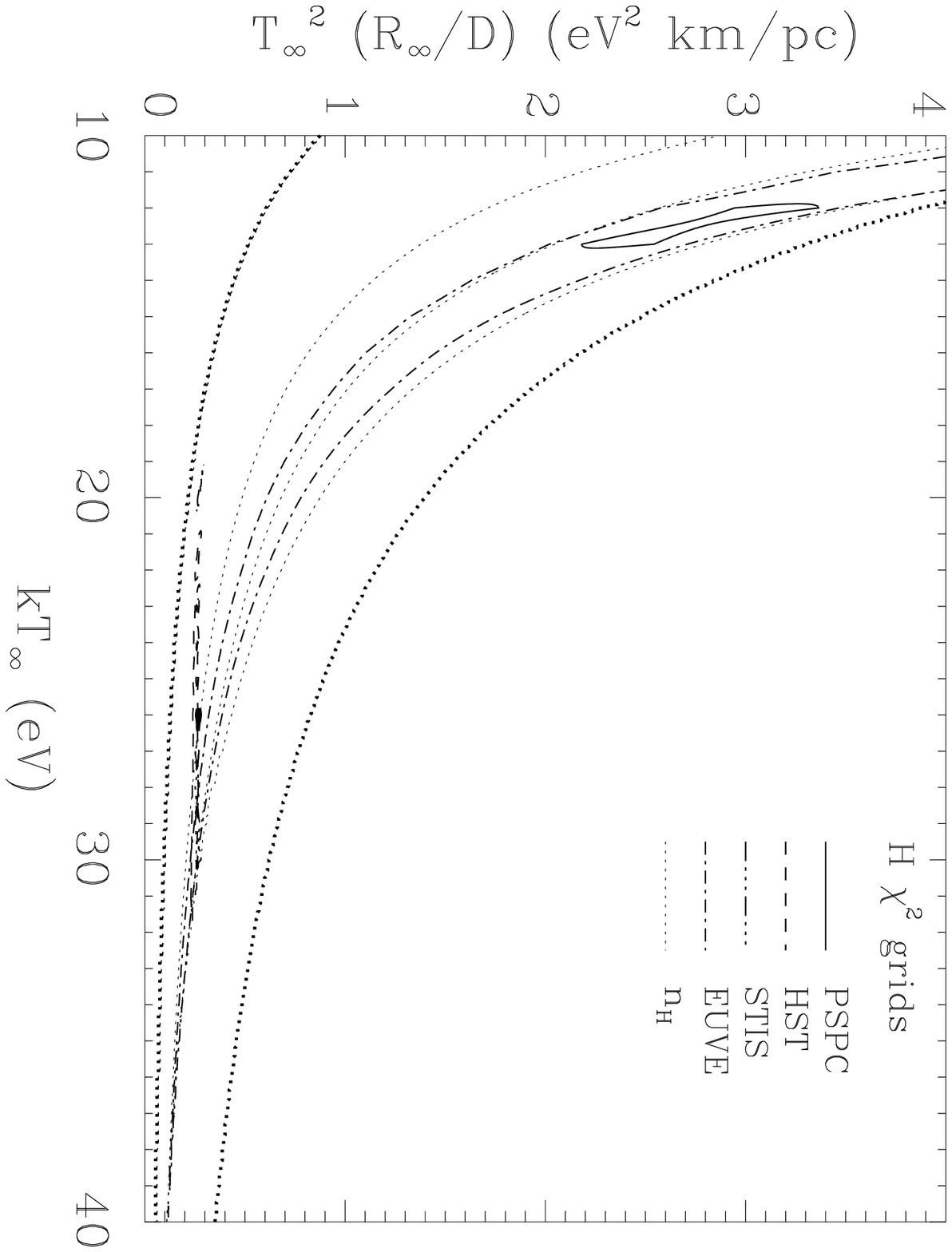}
\end{figure}

\begin{figure}               
\epsscale{0.8}
\plotone{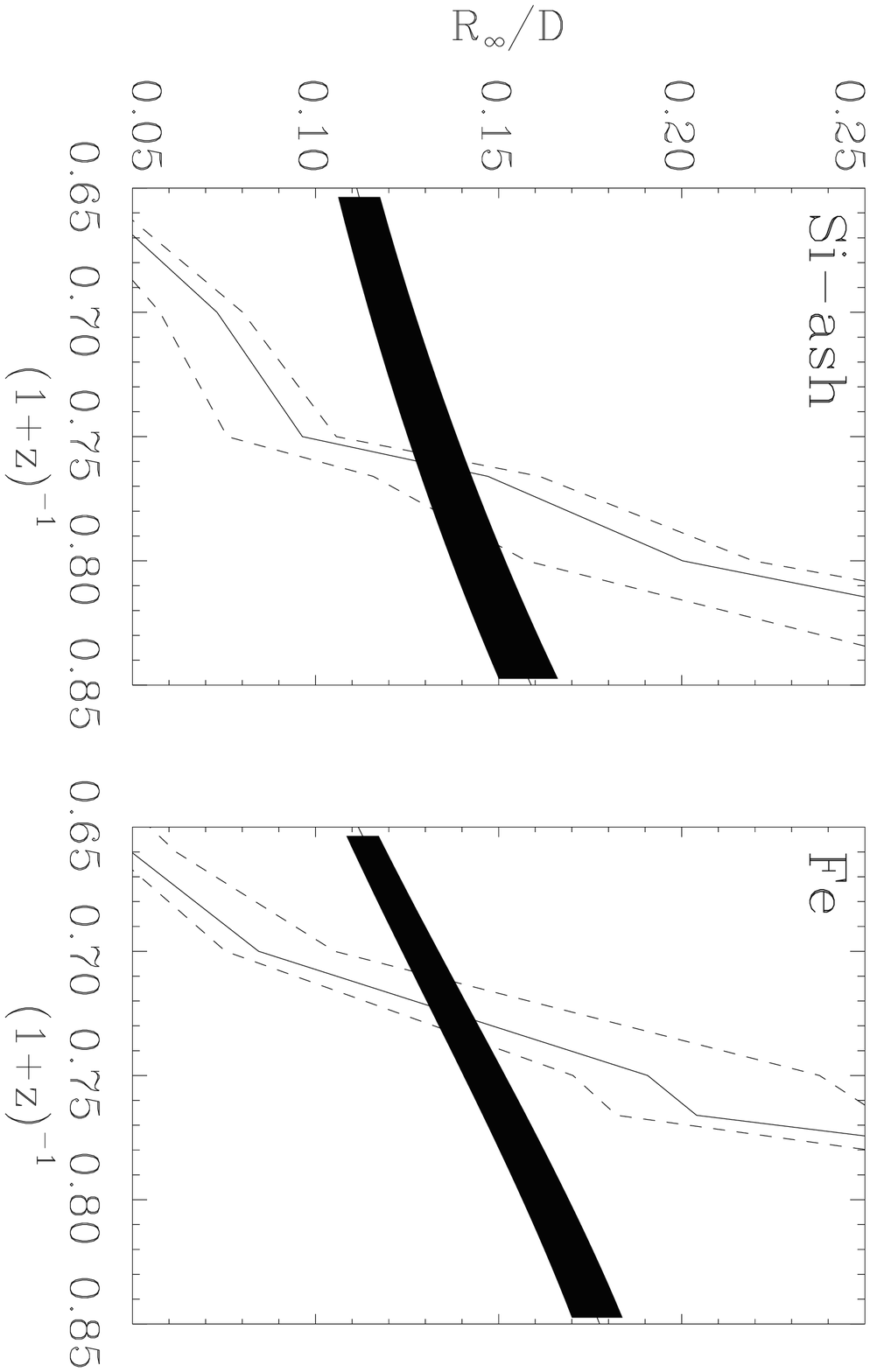}
\end{figure}

\begin{figure}                   
\epsscale{0.8}
\plotone{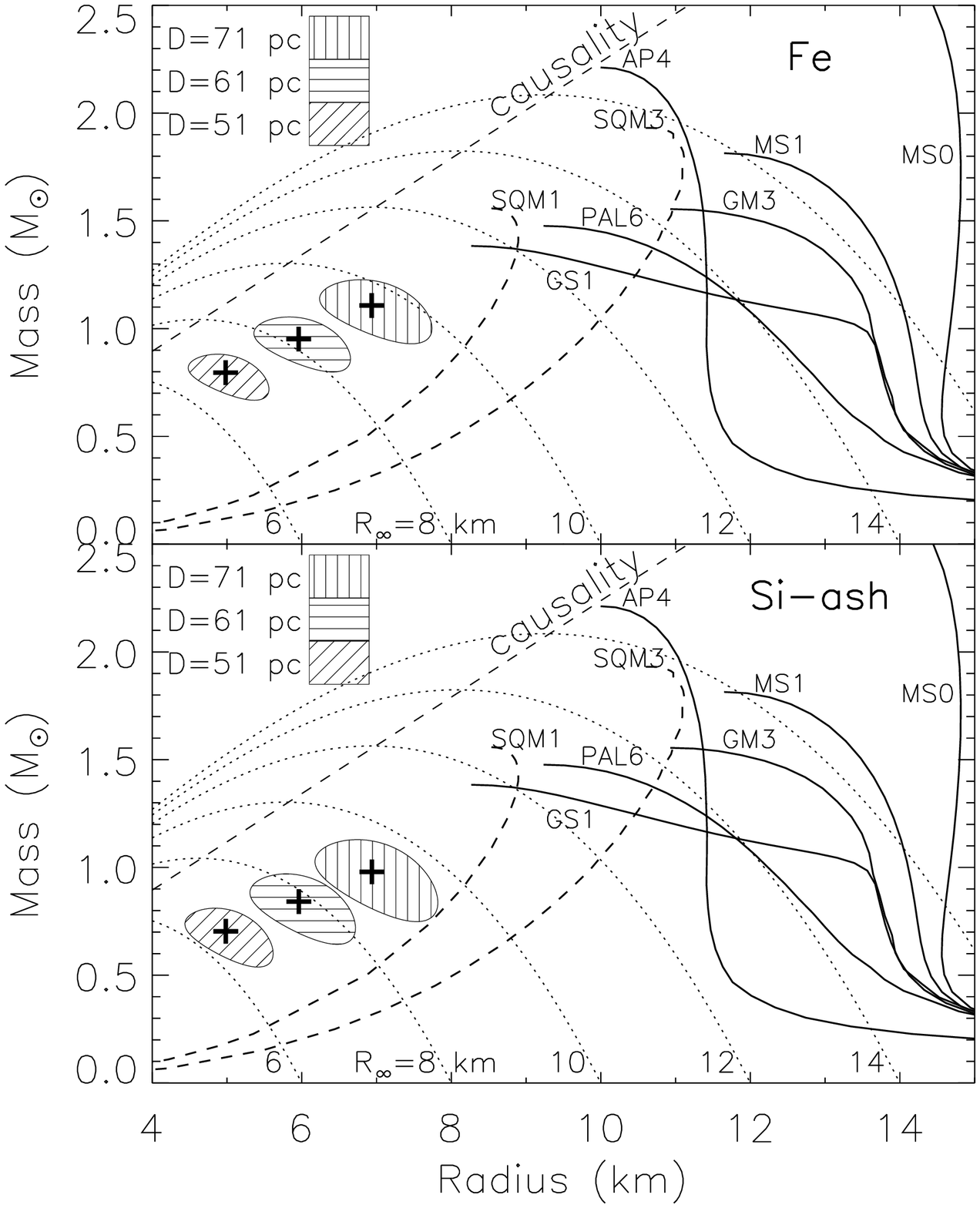}
\end{figure}

\begin{figure}
\epsscale{0.8}
\plotone{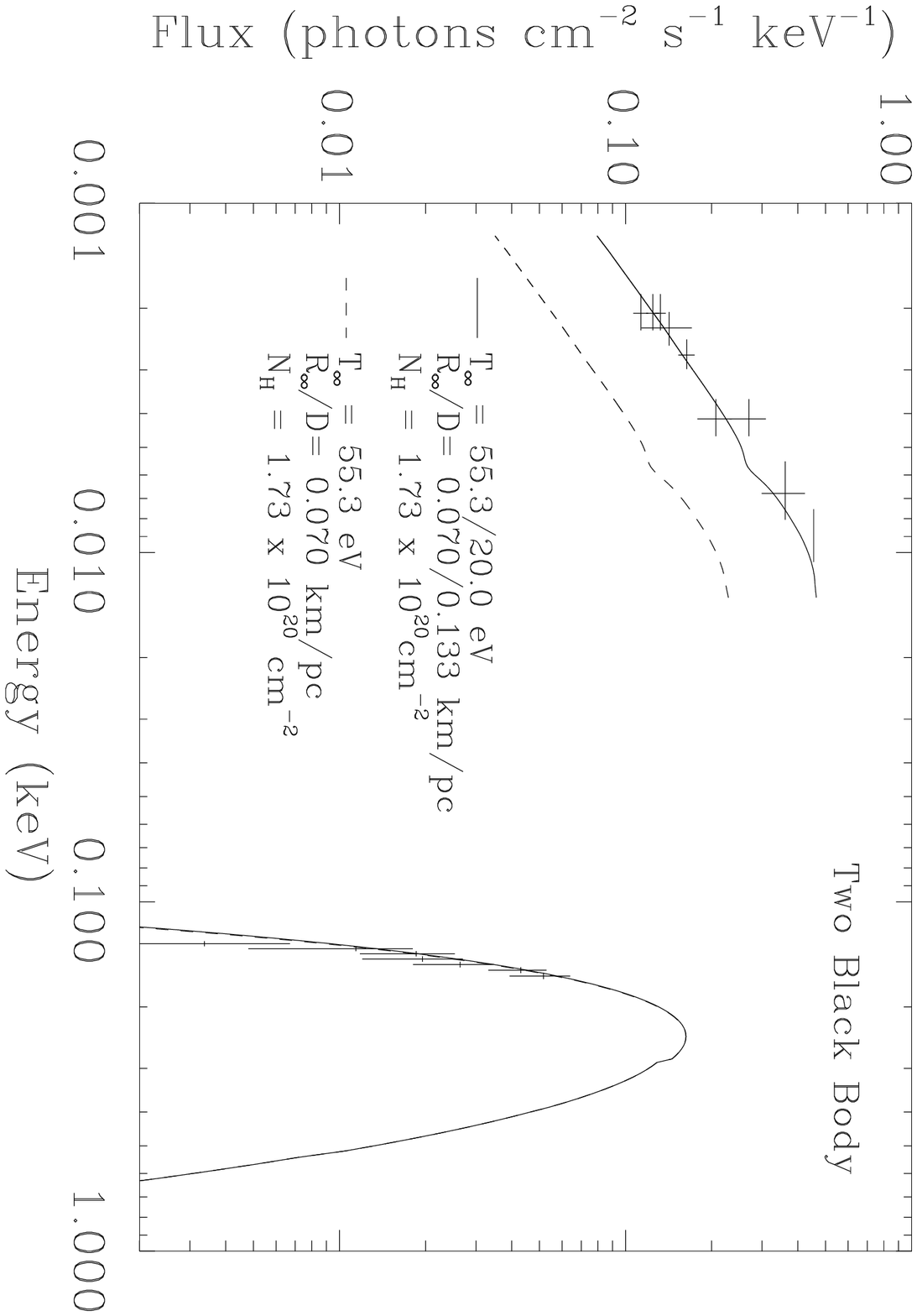}
\end{figure}

\clearpage

\begin{figure}
\epsscale{0.8}
\plotone{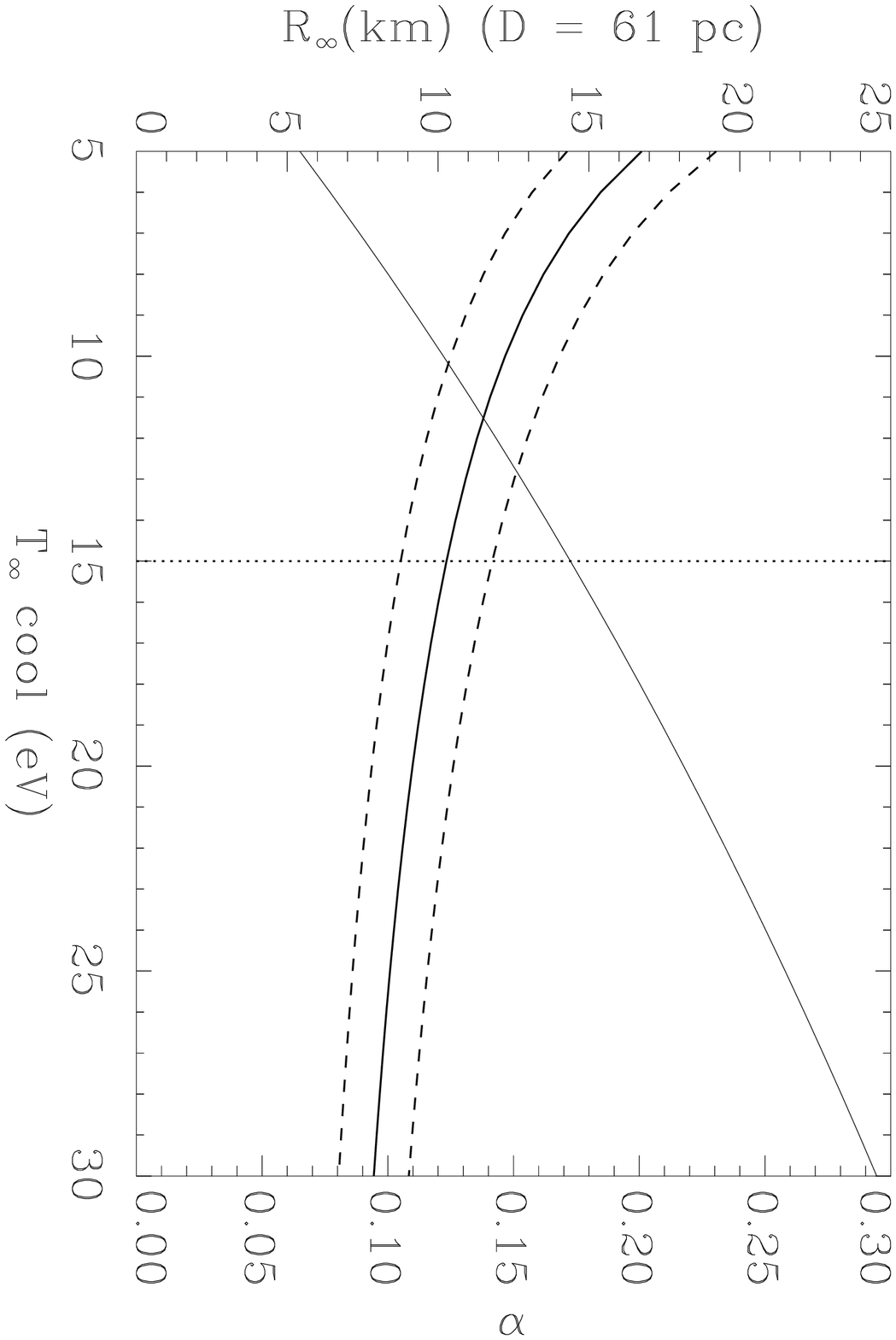}
\end{figure}

\begin{figure}             
\plotone{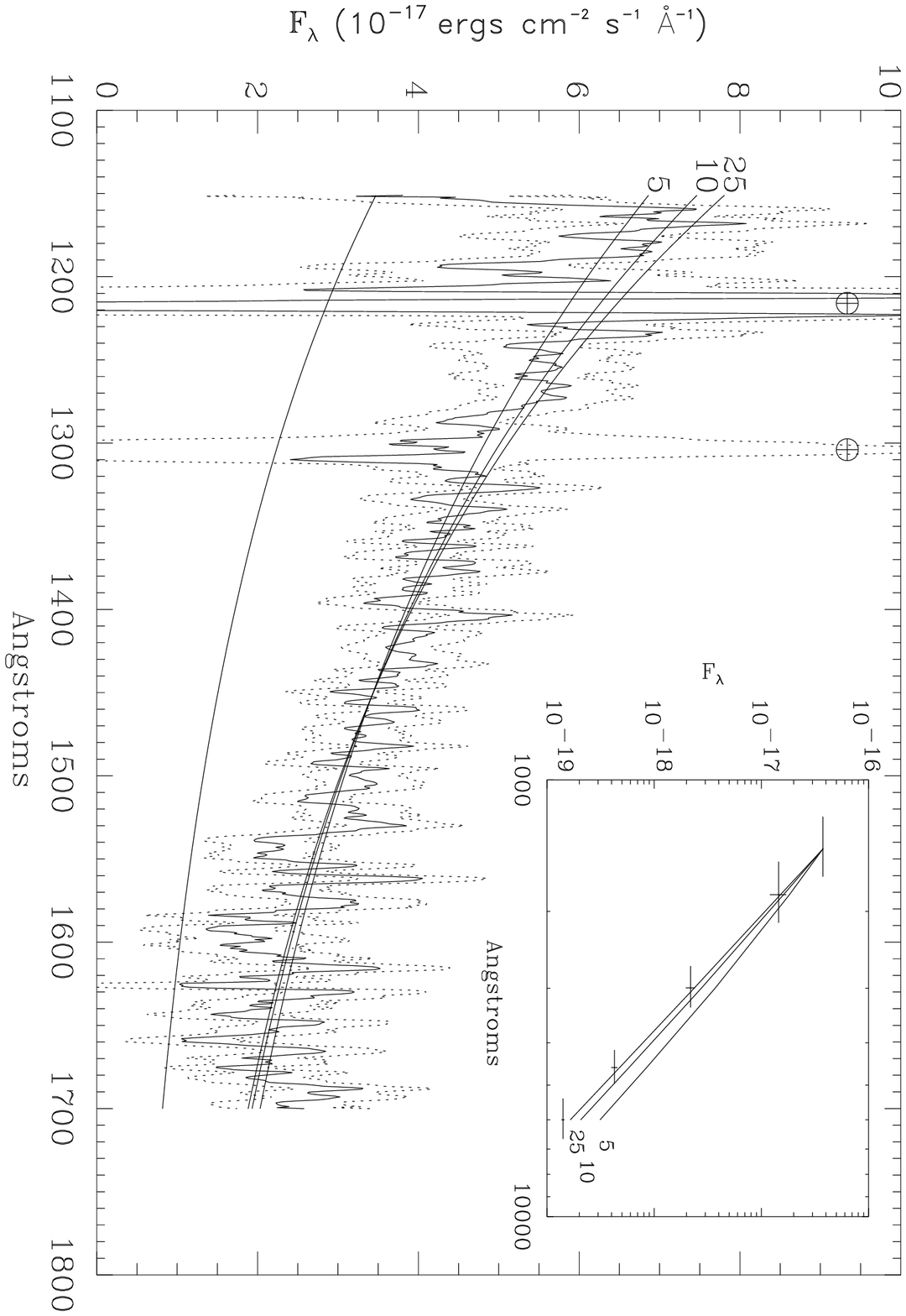}
\end{figure}

\begin{figure}             
\plotone{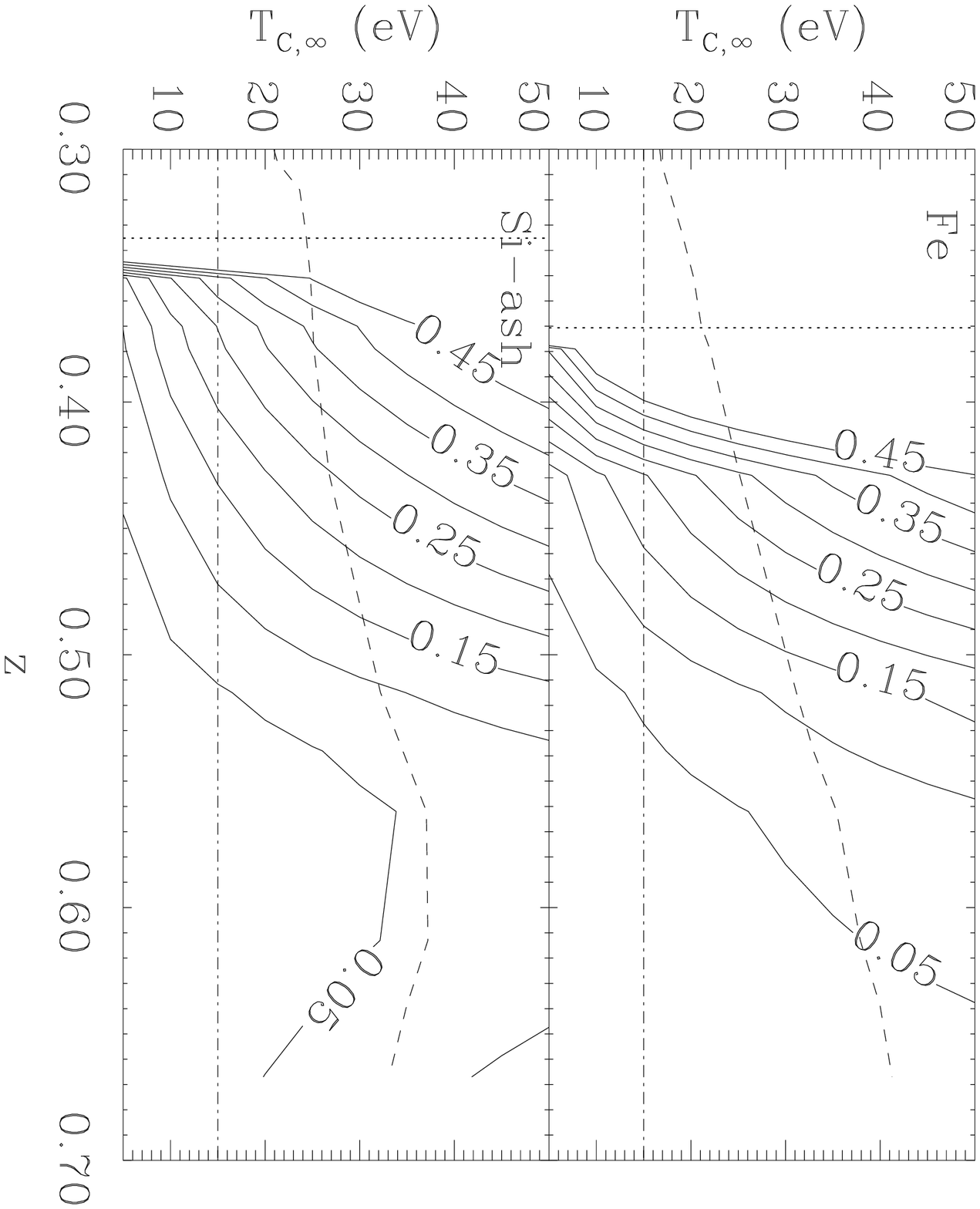}
\end{figure}

\begin{figure}             
\plotone{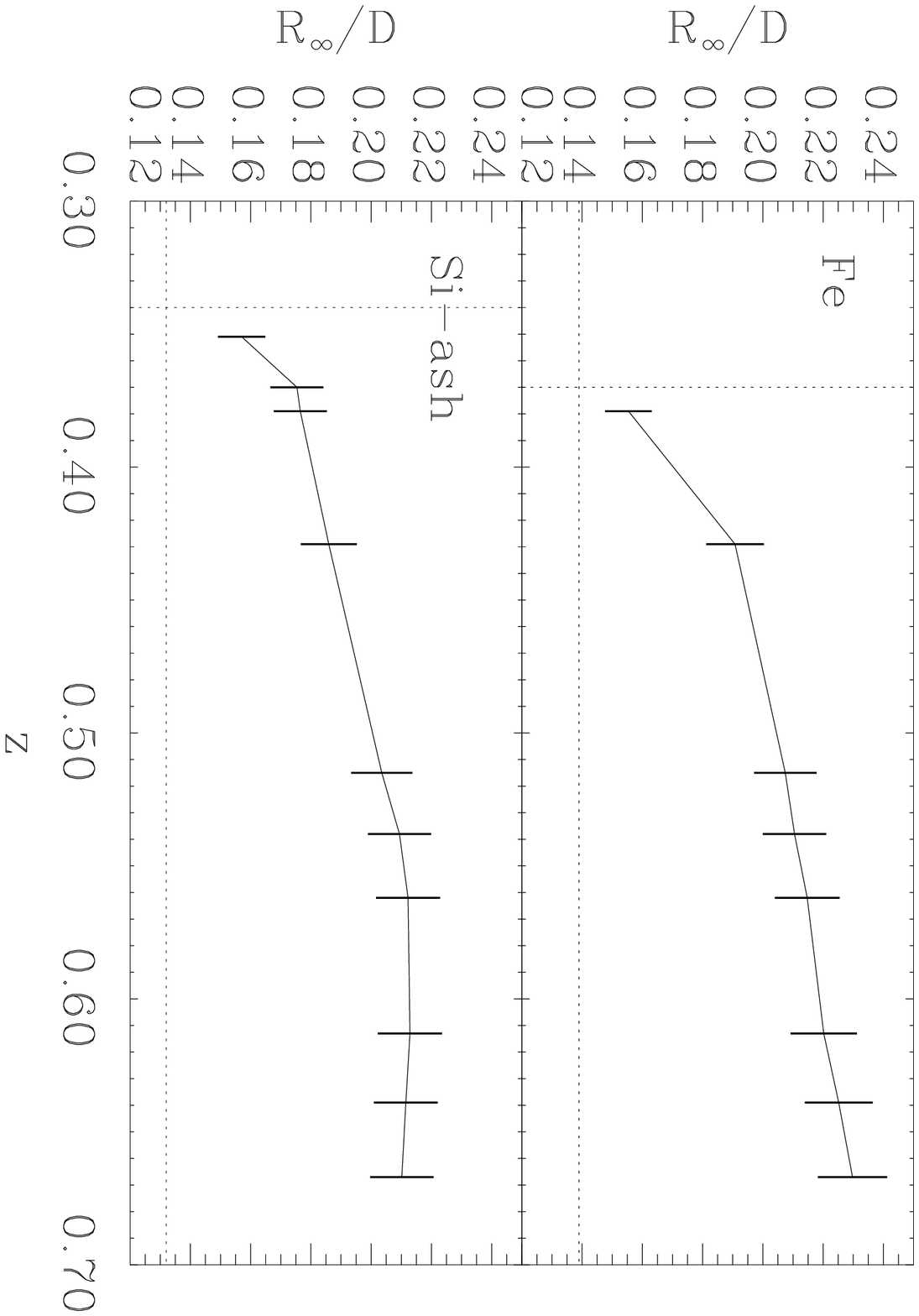}
\end{figure}

\begin{figure}             
\plotone{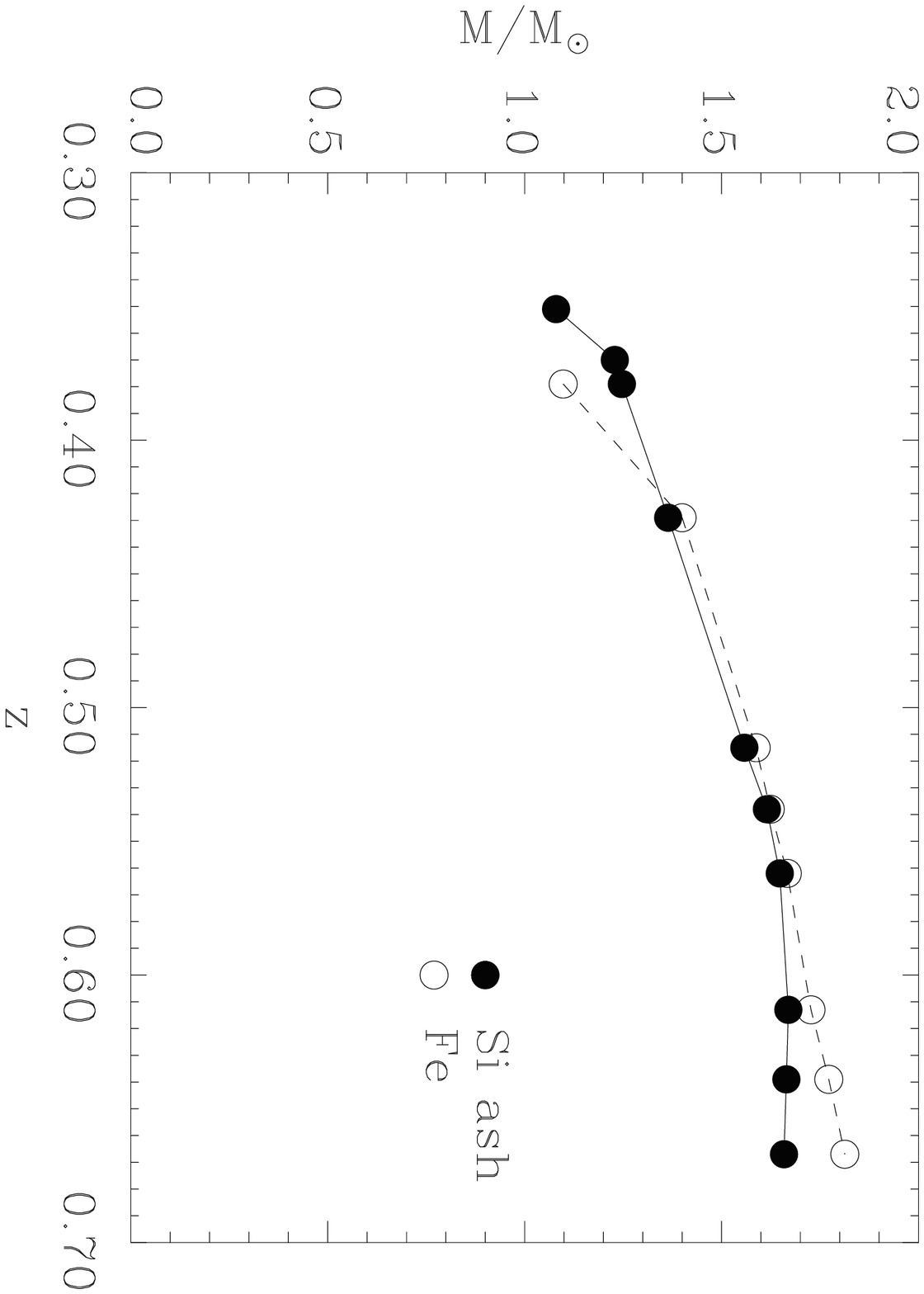}
\end{figure}

\begin{figure}               
\plotone{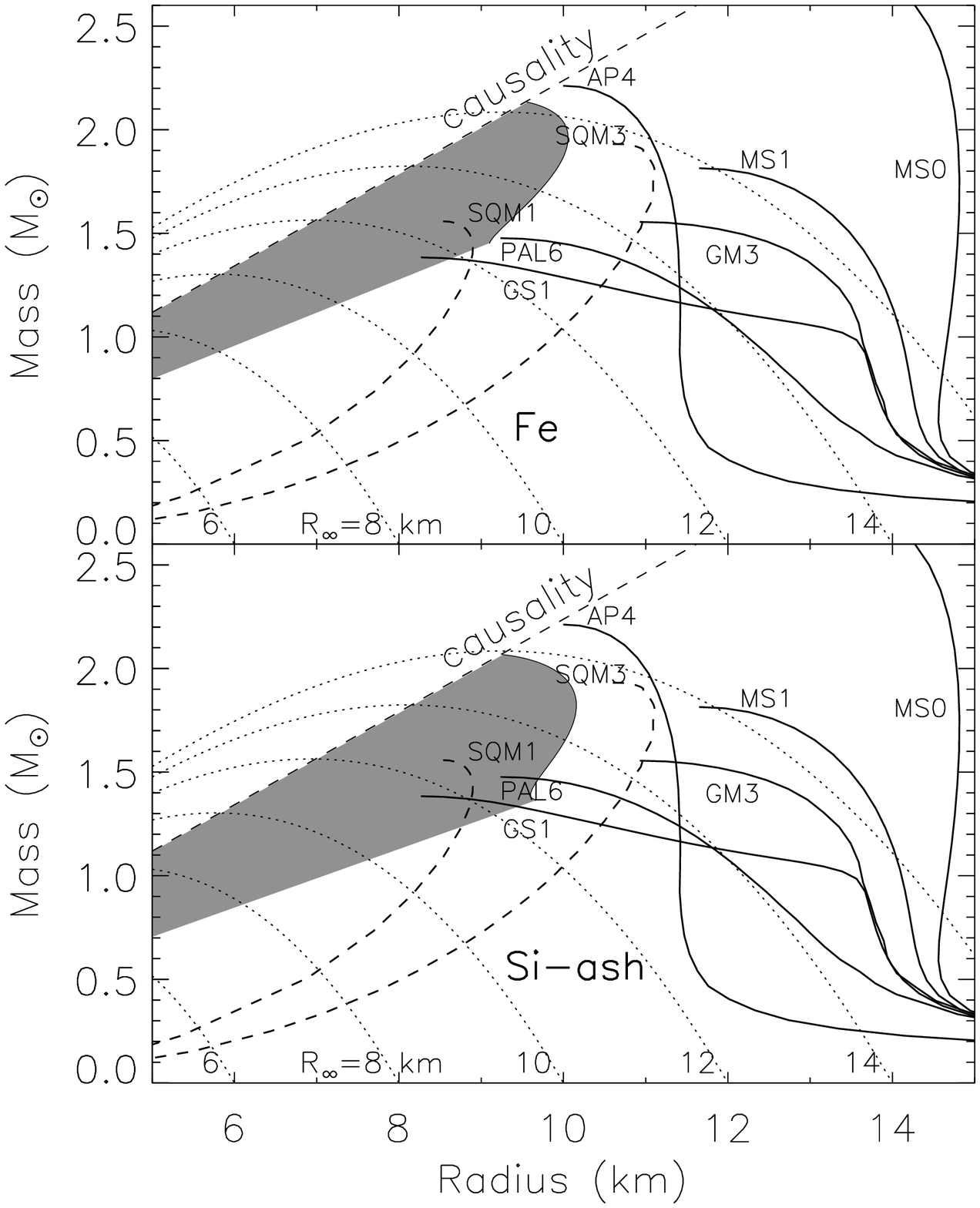}
\end{figure}

\begin{figure}               
\epsscale{0.8}
\plotone{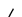}
\end{figure}


\begin{thebibliography}{}

\bibitem[Arnett(1996)]{Arn96}
Arnett, D. 1996, Supernova and Nucleosynthesis (Princeton:Princeton University
                 Press)

 

\bibitem[Caraveo  et al.(1996)]{Car96}
Caraveo, P. A., et al. 1996, \aapr, 7, 209

\bibitem[Christian(1995)]{Chr95}
Christian, D. 1995, ``WSS Memo'' sent to cycle III GOs

\bibitem[Colpi et al.(1998)]{Col98}
Colpi, M., Turolla, R., Zane, S., \& Treves, A. 1998, \apj, 501, 252

\bibitem[Cordes, Romani, \& Lundgren(1993)]{CRL93}
Cordes, J.M., Romani, R.W., \& Lundgren, S.C.  1993, \nat, 362, 133

\bibitem[Dolphin(2000)]{Dol00}
Dolphin, A.E., 2000, \pasp, 112, 1397


\bibitem[Haberl et al.(1997)]{Hab97}
Haberl, F., Motch, C., Buckley, D.A.H., Zickgraf. F.-J., \& Pietsch, W.
1997, \aap, 326, 662 


\bibitem[Haisch, Bowyer \& Malina(1993)]{HBM93}
Haisch, B., Bowyer, S., \& Malina, R. 1993, JBIS, 46, 331

\bibitem[Ho \& Lai(2001)]{HL01}
Ho, W.C.G. \& Lai, D., 2001, \mnras, in press, astro-ph/0104199

\bibitem[Iglesias \& Rogers(1996)]{IR96}
Iglesias, C.A., Rogers, F.J. 1996, \apj, 464, 943


\bibitem[Kulkarni \& Hester(1988)]{KH88}
Kulkarni, S.R. \& Hester, J.J. 1988, \nat, 335, 801

\bibitem[Lampton et al.(1997)]{Lam97}
Lampton, M., Lieu, R., Schmitt, J.H.M.M., Bowyer, S., Voges, W., Lewis, J.,
\& Wu, X. 1997, \apjs, 108, 545

\bibitem[Lattimer \& Prakash(2001)]{LP2000}
Lattimer, J.M. \& Prakash, M., 2001, \apj, 550, 426

\bibitem[Lattimer et al.(1990)]{LPMY}
Lattimer, J.M., Prakash, M., Masak, D. \& Yahil, Y. 1990, \apj, 355, 241 

\bibitem[Madau \& Blaes(1994)]{MB94} 
Madau, R. \& Blaes, O. 1994, \apj, 423, 748 

\bibitem[Mihalas(1978)]{M78}
Mihalas, D. 1978, Stellar Atmospheres (2d ed.; San Francisco:Freeman)

\bibitem[Miller(1992)]{Mil92}
Miller, M.C. 1992, \mnras, 255, 129

\bibitem[Miralles et al.(1993)]{MRL93}
Miralles, J.A., van Riper, K.A., \& Lattimer, J.M. 1993, \apj, 407, 687

\bibitem[Morrison \& McCammon(1983)]{MM83}
Morrison, R. \& McCammon, D. 1983, \apj, 270, 119

\bibitem[Nelson et al.(1995)]{Nel95}
Nelson, R. W., Wang, J. C. L., Salpeter, E. E., \& Wasserman, I. 1995,
\apj, 438, L99

\bibitem[Neuh\"auser, Thomas \& Walter(1998)]{Neu98}
Neuh\"auser, R., Thomas, H.C., \& Walter, F.M. 1998, The Messenger, 92, 27 

\bibitem[\"Ozel (2001)]{Oz01}
\"Ozel, F. 2001, \apj, in press, astro-ph/0103227

\bibitem[Paerels et al.(2000)]{pm2000}
Paerels, F., Mori, K., Motch, C., Haberl, F., Zavlin, V.E., Zane, S., Ramsay,
         G., Cropper, M., \& Brinkman, B. 2000, \aap, 365, L298 

\bibitem[Page(1995)]{Pag95}
Page, D. 1995, \apj, 442, 273
 

\bibitem[Page \& Sarmiento(1996)]{PS96}
Page, D., \& Sarmiento, A.,  1996, \apj, 473, 1067
 
\bibitem[Pavlov et al.(1996)]{Pav96}
Pavlov, G.G., Zavlin, V.E., Truemper, J.  \& Neuh\"auser, R. 1996, 
   \apj, 472, L33 

\bibitem[Prakash, Baron \& Prakash(1990)]{PBP}
Prakash, M., Baron, E. \& Prakash, M. 1990, Phys. Lett., B243, 175

\bibitem[Psaltis, \"Ozel \& DeDeo(2000)]{POD00}
Psaltis, D., \"Ozel, F., \& DeDeo, S., 2000, \apj, 544, 390

\bibitem[Rajagopal \& Romani(1996)]{RR96}
Rajagopal, M. \& Romani, R.W. 1996, \apj, 461, 327

\bibitem[Rajagopal, Romani \& Miller(1997)]{RRM97}
Rajagopal, M., Romani, R.W. \& Miller, M.C. 1997, \apj, 479, 347 


\bibitem[Rutledge et al.(2001)]{rr2001}
Rutledge, R.E., Bildstein, L., Brown, E.F., Pavlov, G.G., \& Zavlin, V.E.
2001, \apj, 551, 921
 
\bibitem[Rogers, Swenson \& Iglesias(1996)]{RSI96}
Rogers, F.J., Swenson, F.J.,  Iglesias, C.A. 1996, \apj, 456, 902

\bibitem[Romani(1987)]{Rom87}
Romani, R.W. 1987, \apj, 313, 718 

\bibitem[Schaab \& Weigel(1998)]{SW98}
Schaab, C. \& Weigel, M.K. 1998, \aap, 336, L13

\bibitem[Seaton(1979)]{Sea79}
Seaton, M.J. 1979, \mnras, 187, P73

\bibitem[Shibanov \& Yakovlev(1996)]{SY96}
Shibanov, Y.A. \& Yakovlev, D.G. 1996, \aap, 309, 171

\bibitem[Thorsett et al.(1994)]{Tho94} 
Thorsett, S.E., Arzoumanian, A., McKinnon, M.M. \& Taylor, H.
1994, \apj, 405, L29 

\bibitem[Thorsett  \& Chakrabarty(1999)]{TC99}
Thorsett, S.E. \& Chakrabarty, D. 1999, \apj, 512, 288


\bibitem[Treves et al.(2000)]{Tre00}
Treves, A., Turolla, R., Zane, S., \& Colpi, M. 2000, \pasp, 112, 297

\bibitem[van Kerkwijk \& Kulkarni(2000)]{KK00}
van Kerkwijk, M. \& Kulkarni, S. 2000, ESO press release PR 19/00

\bibitem[Walter(2001)]{Wal01}
Walter, F.M. 2001, \apj, 549, 433

\bibitem[Walter \& Matthews(1997)]{WM97}
Walter, F.M. \& Matthews, L.D. 1997, \nat, 389, 358

\bibitem[Walter \& Wijers(2001)]{bows}
Walter, F.M. \& Wijers, R.A.M.J. 2001, in preparation.

\bibitem[Walter, Wolk \&  Neuh\"auser(1996)]{WWN96}
Walter, F.M., Wolk, S.J., \& Neuh\"auser, R. 1996, \nat, 379, 233

\bibitem[Wanajo et al.(2001)]{WKMO01}
Wanajo, S., Kajino, T., Matthews, G.J., \& Otsuki, K. 2001, \apj, 554, 578

\bibitem[Wang et al.(1999)]{Wan99}
Wang, J.C.L., Link, B., van Riper, K., Arnaud, K.A., \& Miralles, J.A.
1999, \aap, 345, 869


\bibitem[Witten(1984)]{witten}
Witten, E. 1984, \prd, D30, 272 

\bibitem[Woodgate et al.(1998)]{Woo98}
Woodgate, B.E., et al. 1998, \pasp 110, 1183

\bibitem[Zampieri et al.(1995)]{Zam95}
Zampieri, L., Turolla, R., Zane, S., \& Treves, A. 1995, \apj, 439, 849

\bibitem[Zane, Turolla \& Treves(2000)]{ZTT00} 
Zane, S., Turolla, R., \& Treves, A., 2000, \apj, 537, 387

\bibitem[Zavlin et al.(1995)]{Zav95}
Zavlin, V.E., et al. 1995, \aap, 297, 441

\bibitem[Zavlin, Pavlov \& Shibanov(1996)]{ZPS96}
Zavlin, V.E., Pavlov, G.G., Shibanov, Yu.A. 1996, \aap, 315, 141

\end{thebibliography}
\end{document}